\documentclass[12pt]{article}
\newcommand{\acks}[1]{\section*{Acknowledgments}#1}
\usepackage{blindtext}
\usepackage[round]{natbib}
\usepackage{amsthm}
\usepackage{amssymb}
\usepackage[colorlinks=true,
            linkcolor=blue,
            citecolor=blue,
            urlcolor=blue]{hyperref}

\usepackage[linesnumbered,ruled,vlined]{algorithm2e}
\SetKwInput{KwInput}{Input}              \SetKwInput{KwOutput}{Output}

\usepackage{array}
\usepackage[caption=false,font=normalsize,labelfont=sf,textfont=sf]{subfig}
\usepackage{textcomp}
\usepackage{stfloats}
\usepackage{url}
\usepackage{verbatim}
\usepackage{bm}
\usepackage{booktabs}
\usepackage{bbm}
\usepackage{mathrsfs}
\usepackage{amsfonts}
\usepackage{amsmath}
\usepackage{graphicx}

\usepackage{placeins}
\usepackage{rotating}

\usepackage{xcolor}

\usepackage[normalem]{ulem}
\newcommand{\blind}{1}

\usepackage{multirow}
\usepackage{lscape}

\usepackage[noend]{algpseudocode}

\makeatletter
\def\BState{\State\hskip-\ALG@thistlm}
\makeatother

\usepackage{enumitem}



\newtheorem{lemma}{Lemma}
\newtheorem{proposition}{Proposition}
\newtheorem{corollary}{Corollary}
\newtheorem{remark}{Remark}

\usepackage{lastpage}

\usepackage{lastpage}

\usepackage{lineno}
\usepackage{setspace}
\usepackage[margin=0.7in]{geometry}
\if1\blind
{
\title{MAPLE: Mapper Based Localized Prediction with Data Driven Cover Selection for High dimensional Data}

\author{
Md Moinul Ahsan$^{1}$ \and
Priyam Das$^{1}$ \and
Nitai D Mukhopadhyay$^{1}$ \\[0.5em]
$^{1}$Department of Biostatistics, Virginia Commonwealth University, \\
Richmond, VA, USA \\
}

} \fi

\if0\blind
{

\title{MAPLE: Mapper Based Localized Prediction with Data Driven Cover Selection for High dimensional Data}

} \fi

\date{}

\definecolor{vividgreen}{HTML}{1A9850}
\definecolor{lightgreen}{HTML}{ADDD8E}
\definecolor{yellowish}{HTML}{FEE10B}
\definecolor{redish}{HTML}{F46D43}
\definecolor{darkredish}{HTML}{A50026}

\usepackage{makecell}

\begin{document}
\maketitle
\begin{abstract}
High dimensional biomedical data often exhibit nonlinear, heterogeneous, and manifold driven structures that challenge global parametric and tree-based models. We propose MAPLE (mapper-based Adaptive Prediction via Local Estimation), a localized prediction framework grounded in topological data analysis. The method is formulated as a nonparametric estimator of conditional class probabilities that adapts to the intrinsic geometry of the predictor space. Neighborhoods are defined through connectivity in a data-adaptive Mapper graph, enabling localized averaging within graph induced regions that capture complex structures such as branching and multi-scale heterogeneity. We introduce a statistically principled, data driven procedure for cover selection based on a bias–variance trade off, yielding optimal asymptotic scaling for interval widths and overlaps. The framework accommodates binary, nominal, and ordinal outcomes and incorporates a permutation-based variable importance measure to quantify covariate contributions in prediction. We establish theoretical guarantees, including pointwise consistency and Bayes risk consistency under standard regularity conditions. Simulations show that MAPLE consistently outperforms or matches multinomial regression, ordinal regression, and random forest, with the largest gains observed under heterogeneous and high-noise settings. Applications to Parkinson’s disease progression (PPMI) and glioma classification (TCGA RNA sequencing) demonstrate strong predictive accuracy and interpretable, topology-aware summaries of underlying data structure.
\end{abstract}


\clearpage
\section{Introduction}\label{sec:intro}

High dimensional data are increasingly encountered in modern biomedical and clinical research, including genomics, neuroimaging, and electronic health records. In many such applications, the relationship between predictors and outcomes is characterized by substantial nonlinearity, heterogeneity, and latent structure driven by underlying biological processes such as disease progression, cellular differentiation, or molecular subtyping. Although measured in high dimensional spaces, these data are often well approximated by low dimensional manifolds, reflecting the intrinsically smooth and constrained nature of these biological processes \citep{cunningham2014dimensionality, trapnell2014dynamics}. This structure poses significant challenges for predictive modeling, particularly when predictor response relationships vary across subpopulations or exhibit localized patterns that are not well captured by global models such as multinomial or proportional odds regression models.

Classical regression approaches, including multinomial and proportional odds models, rely on global functional assumptions that may be severely misspecified in the presence of nonlinearity and heterogeneity. Nonparametric methods, such as nearest neighbor and kernel based estimators, address this limitation by performing localized averaging \citep{cover1967nearest, nadaraya1964estimating, devroye2013probabilistic}. However, these approaches define neighborhoods through fixed metric notions of proximity or bandwidth parameters, which may fail to adapt to the intrinsic geometry of high dimensional data, particularly when observations lie on complex manifolds or exhibit branching and multi scale structure. Tree based and ensemble methods, such as random forest, provide flexible alternatives through adaptive partitioning of the predictor space \citep{breiman2001random, hastie2009elements}, but they may not explicitly incorporate geometric or connectivity information and thus may offer limited interpretability in structured settings. Model averaging and ensemble based approaches have also been proposed, including boosting based methods and distribution-free smoothing techniques for combining multi category diagnostic information \citep{li2022adaboost, maiti2023distribution}. Nevertheless, these approaches primarily operate through aggregation of model outputs and do not explicitly leverage the geometric structure of the data.

Topological data analysis (TDA) provides a complementary framework for analyzing complex data by characterizing their geometric and connectivity structure \citep{carlsson2009topology, wasserman2018topological, chazal2021introduction}. Among TDA tools, the Mapper algorithm \citep{singh2007topological} constructs a graph based representation of the data by combining a filter function, an overlapping cover of the filter space, and clustering within induced subsets. This representation captures key structural features such as branching, connectivity, and multi scale heterogeneity that are often difficult to recover using classical statistical methods. Mapper has been successfully applied in exploratory analyses of biomedical data, including the identification of clinically meaningful subgroups and disease phenotypes \citep{nicolau2011topology, lum2013extracting}. Despite its empirical success, Mapper has seen limited development as a statistically principled framework for supervised learning and estimation. Existing approaches are largely exploratory or heuristic and lack a formal connection to statistical learning theory. While recent work has established structural stability and structural properties of the one dimensional Mapper construction \citep{carriere2018structure}, these results do not address several fundamental challenges in its role as a predictive estimator. First, the construction of the Mapper graph depends on heuristically selected tuning parameters, without a principled bias–variance justification. Second, Mapper does not incorporate response data into its model and, therefore, cannot define an explicit estimator of the predictive relationship. Third, supervised extensions of Mapper typically do not produce estimators with established statistical guaranties.

Our contributions are threefold. First, we propose Mapper-based Adaptive Prediction via Local Estimation (MAPLE) within the framework of nonparametric regression and classification by defining an explicit estimator of conditional class probabilities. Second, we introduce a principled, data driven procedure for selecting cover parameters based on a bias-variance trade off, extending classical smoothing theory to the Mapper setting. Finally, we establish theoretical guarantees for the proposed estimator, including pointwise consistency and Bayes risk consistency under regularity conditions. In addition, the proposed framework accommodates binary, nominal, and ordinal outcomes and includes a permutation based variable importance measure for assessing covariate contributions to predictive performance.

The remainder of the paper is organized as follows. Section \ref{sec2} presents the proposed methodology and establishes its theoretical properties. Section \ref{sec3} provides a comprehensive simulation study designed to evaluate performance under nonlinear, heterogeneous, and high dimensional settings, incorporating features such as branching structures, skewed covariates, correlated predictors, and the presence of irrelevant or omitted variables, including additional scenarios to assess robustness and variable selection. Section \ref{sec4} demonstrates the practical utility of the proposed method through applications to two biomedical datasets: prediction of disease stage in Parkinson’s disease using data from the Parkinson’s Progression Markers Initiative (PPMI), and classification of tumor severity using RNA sequencing data from The Cancer Genome Atlas (TCGA). Section \ref{sec5} concludes with a discussion.

\section{Method}\label{sec2}

Let ${(x_i, y_i)}_{i=1}^n$ denote independent observations, where $x_i \in \mathbb{R}^p$ is a $p$ dimensional predictor vector and $y_i \in \{1, \ldots, R\}$ is an ordinal response with $R$ ordered classes. We define a one dimensional filter variable $T = f(X)$, obtained from the predictor vector through a specified mapping $f:\mathbb{R}^p\to\mathbb{R}$. Our objective is to estimate the conditional class probabilities
\[
\eta_r(t) = \mathbb{P}(Y = r \mid T = t), \quad r = 1,\ldots,R,
\]
and to construct a prediction rule for a new observation $x_{\mathrm{new}}$ through its induced filter value $t_{\mathrm{new}} = f(x_{\mathrm{new}})$. Thus, the problem is formulated as nonparametric estimation in the one dimensional space induced by the filter $T$, while neighborhood structure is learned from the geometry of the original predictor space through the Mapper graph. Although the estimator is indexed by $t$, it is constructed using graph based neighborhoods defined in the predictor space, which induce local neighborhoods in the filter space under the regularity conditions described below. We begin by briefly reviewing the standard Mapper construction before describing the proposed method.

\subsection{Background on the Mapper Algorithm}

The Mapper algorithm was introduced by \cite{singh2007topological} primarily as a tool for visualization and exploratory analysis of high dimensional data through a combination of dimensionality reduction, clustering, and graph construction. It produces a graph based representation that captures the geometric and connectivity structure of the data \citep{vannoni2024exploitation}. The standard Mapper construction proceeds in three steps. Firstly, a filter function (or lens) $f : X \rightarrow \mathbb{R}^d$ maps the data from a high dimensional space into a lower dimensional space, typically with $d=1$. The choice of filter determines which structural features of the data are emphasized and is often guided by the application \citep{chazal2021introduction, lum2013extracting}. Secondly, the range of the filter function is partitioned into $l$ overlapping intervals $\{I_j\}_{j=1}^l$. These intervals induce subsets of the original data through inverse images,
\[
U_j = \{ x_i \in X : f(x_i) \in I_j \}, \quad j=1,\dots,l.
\]
The overlap between intervals ensures that neighboring subsets share observations, enabling the representation to capture connectivity across the data. The number of intervals and the degree of overlap are tuning parameters that control the resolution of the construction.
Thirdly, within each subset $U_j$, a clustering algorithm is applied to obtain clusters $\{C_{j,k}\}$. Each cluster corresponds to a node in the Mapper graph. Edges are introduced between nodes whenever clusters from overlapping (or adjacent) intervals share at least one observation. The resulting graph encodes both local grouping and global connectivity of the data.

\begin{figure}[ht]
\centering
\includegraphics[width=0.99\textwidth]{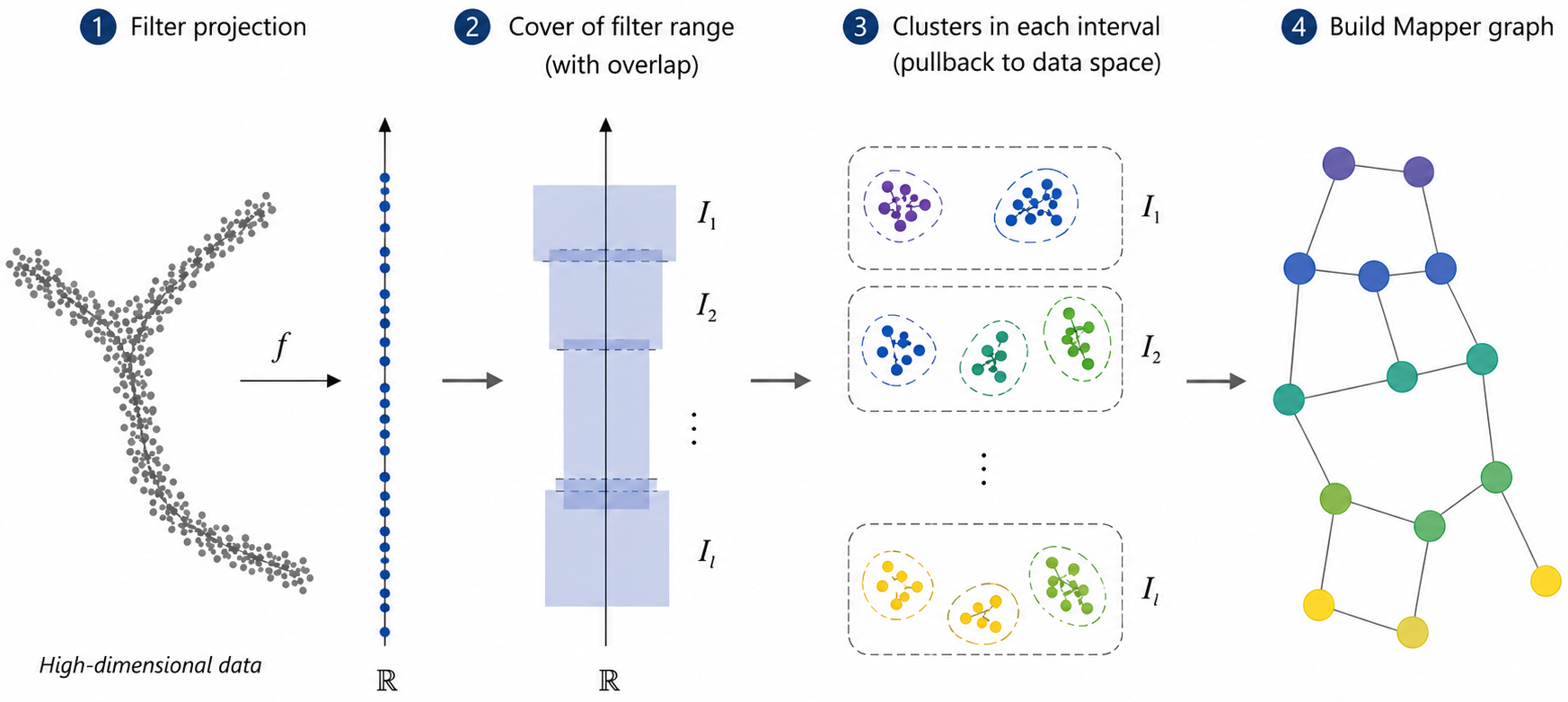}
\caption{
Illustration of the Mapper construction. High dimensional observations are projected onto a one dimensional filter function, whose range is partitioned into overlapping intervals. Clustering within the corresponding pullback subsets produces nodes in the Mapper graph, while shared observations across overlapping intervals induce graph connectivity. The resulting representation summarizes both local grouping and global geometric structure.
}
\label{fig:mapper_workflow}
\end{figure}

\subsection{Proposed MAPLE Construction}

In our implementation, we define a one dimensional filter function $f : \mathbb{R}^p \to \mathbb{R}$ using Ordinal Accelerated Sparse Discriminant Analysis (OASDA; \citealp[]{han2024sparse}), given by
\[
f(x_i) = \mathbf{w}^\top x_i,
\]
where $\mathbf{w} \in \mathbb{R}^p$ is a projection vector estimated through a penalized optimization problem that incorporates both ordinal structure in the response and sparsity regularization. The OASDA framework employs an accelerated proximal gradient algorithm, ensuring stable estimation in high dimensional settings ($p \gg n$) \citep{cardoso2007learning, clemmensen2011sparse, han2024sparse}. We consider Mapper construction using a unidimensional filter function, which remains both convenient and mainstream, with well-established structural and stability guarantees \citep{carriere2018structure}. The resulting one dimensional projection is denoted by as $t_i = f(x_i)$. The range of $\{t_i\}_{i=1}^n$ is then partitioned into $l$ overlapping intervals $\{I_j\}_{j=1}^{l}$, each defining a subset
\[
U_j = \{ x_i : f(x_i) \in I_j \}.
\]
Within each subset, we perform hierarchical clustering using Ward’s criterion based on Euclidean distance \citep{ward1963hierarchical}, generating clusters $\{C_{j,k}\}_{k=1}^{K_j}$, where the number of clusters $K_j$ is selected using the silhouette criterion \citep{rousseeuw1987silhouettes}. For each fixed interval \(j\), the clusters \(\{C_{j,k}\}_{k=1}^{K_j}\) form a disjoint partition of \(U_j\). However, because adjacent Mapper intervals overlap, clusters from different intervals may share observations. Such shared observations define the edges of the Mapper graph $\mathcal{G}_n$, yielding a data adaptive partition of the predictor space that reflects its intrinsic geometry.

\subsection{Data Driven Cover Selection}

MAPLE predictions depend on the resolution of the cover in filter space; therefore, the number of intervals and their overlap must be chosen adaptively. We adopt a bias-variance framework to select the cover for the one dimensional filter \(T=f(X)\). Let \(l\) denote the number of intervals, \(S\) the number of ordered observations contained in each interval, and \(q\) the number of ordered observations by which consecutive intervals are shifted along the ordered filter values \(t_{(1)} \le \cdots \le t_{(n)}\). Thus, if one interval begins at \(t_{(i)}\), the next interval begins at \(t_{(i+q)}\), implying an overlap of \(S-q\) observations between adjacent intervals. The \(j\)-th interval is defined by 
\[
I_j = \left\{ 
t_{((j-1)q+1)}, \ldots, t_{((j-1)q+S)} 
\right\},
\qquad j = 1, \ldots, l.
\]
The total sample size therefore satisfies $S + (l-1)q = n$. This construction yields intervals with approximately equal sample sizes, providing stable local estimation even in highly skewed regions of the filter distribution where fixed-width intervals may contain very few observations. Smaller values of \(q\) increase overlap, whereas larger values reduce overlap. However, excessive overlap causes observations to contribute repeatedly across neighboring Mapper nodes, increasing redundancy and potentially oversmoothing localized structure in the filter space. To limit excessive redundancy, we restrict each observation to belong to at most two consecutive intervals. This condition is equivalent to requiring that nonadjacent intervals do not overlap. Specifically, triple membership occurs if and only if $I_j \cap I_{j+2} \neq \emptyset$, which requires $(j-1)q + S \ge (j+1)q + 1$. Rearranging gives $S \ge 2q+1$, or equivalently $S>2q$. Therefore, a necessary and sufficient condition to ensure that each observation belongs to at most two consecutive intervals is $q \ge \frac{S}{2}$. In addition, $q>S$ produces disjoint intervals with no overlap between adjacent subsets. Hence, the admissible range ensuring overlap only between adjacent intervals is $\frac{S}{2} \le q \le S$.
\begin{figure}
    \centering
    \includegraphics[width=1\linewidth,height=0.35\textheight]{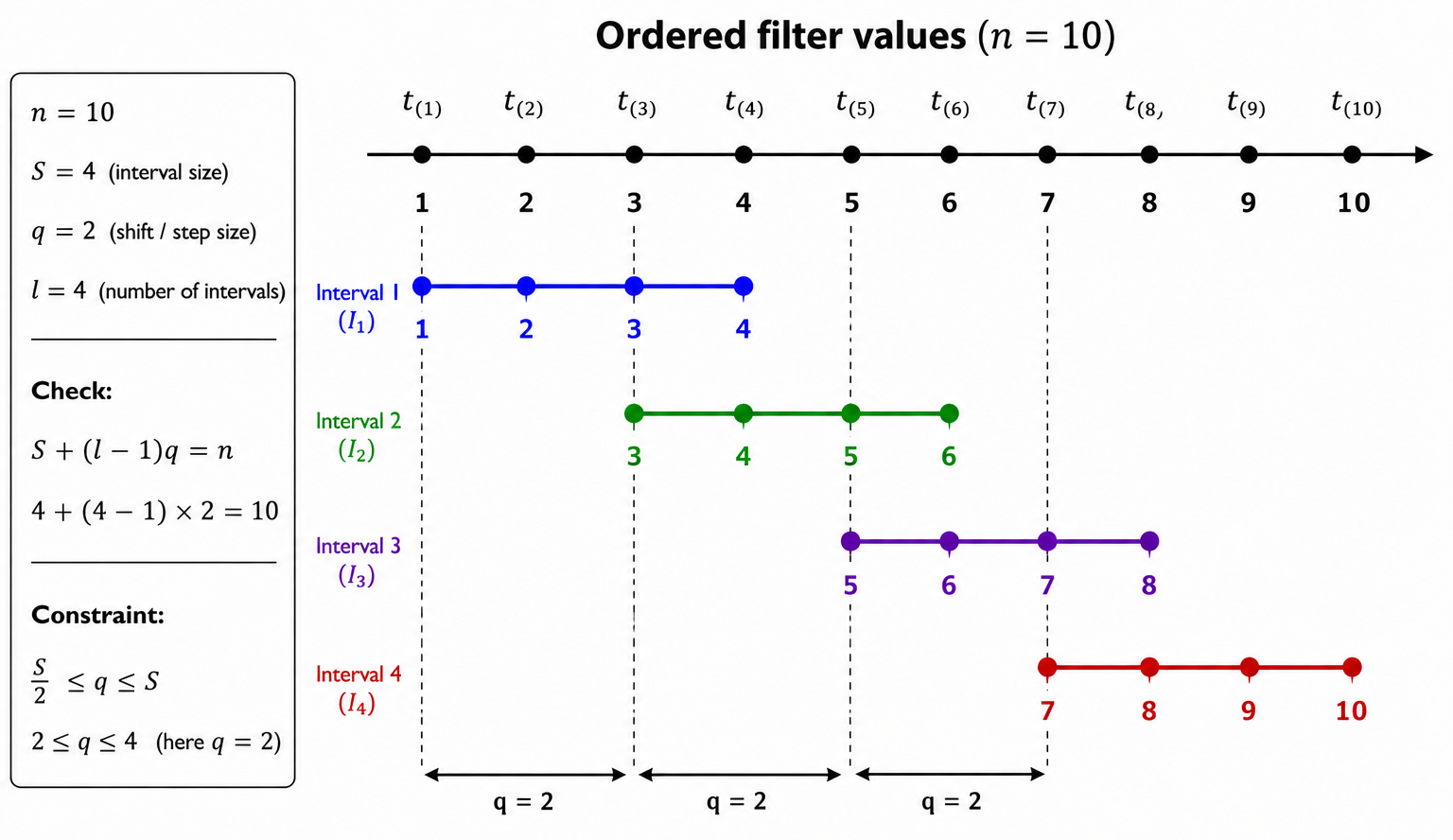}
    \caption{Overlapping interval construction in ordered filter space with interval size \(S\), shift size \(q\), and \(l\) intervals satisfying the constraint.}
    \label{cover_selection}
\end{figure}
Figure \ref{cover_selection} provides an illustrative example of the overlapping cover construction for \(n=10\) ordered filter values, where each interval contains \(S=4\) observations and consecutive intervals are shifted forward by \(q=2\) ordered observations. Each new interval therefore overlaps the previous interval by two observations while contributing two new observations to the cover.

Increasing \(l\) produces a finer cover that better captures local structure but increases variability, whereas decreasing \(l\) stabilizes estimation at the cost of over smoothing. Under standard smoothness conditions on $\eta_r(t)$, the variance scales as \(S^{-1}\) and the squared bias scales as \(l^{-4}\), consistent with classical nonparametric regression theory \citep{wasserman2006all, freedman1981histogram} and our derivation (Supplementary Section S1). This motivates the criterion
\begin{equation}
\label{eq:cover_loss}
L(S,l)=\rho S^{-1}+(1-\rho)l^{-4}, \quad \rho\in(0,1).
\end{equation}

\begin{lemma}[\textit{Optimal cover scaling}]
Let $(S^*,l^*)$ minimize \eqref{eq:cover_loss} subject to the constraint $S+(l-1)q=n$, where $q$ is determined implicitly by the constraint. Then
\[
l^* = \left( \frac{8n(1 - \rho)}{\rho} \right)^{1/5}, \qquad 
S^* = \frac{2n}{l^*+1}, \qquad 
q^* = \frac{n}{l^*+1},
\]
and the optimal overlap ratio satisfies $q^*/S^* = 1/2$.
\end{lemma}
In practice, the optimal values $(S^*,l^*,q^*)$ may not be integers. We therefore construct a feasible cover by discretizing these quantities, 
taking integer values of $(l^*,S^*)$, and determining $q$ from the constraint
\[
S + (l-1)q = n \quad \Rightarrow \quad (l-1)q = n - S
\quad \Rightarrow \quad
q = \frac{n - S}{l-1}.
\]
This choice ensures that the resulting intervals cover the full range of the filter values while maintaining the prescribed level of overlap. Substituting the optimal value $S^*$ into the constraint yields the expression for $q^*$ in the lemma (see Supplementary Section~S1). The tuning parameter $\rho$ is selected via cross validation over a grid $\rho \in \{0.1,\dots,0.9\}$, choosing the value that minimizes out of sample prediction error \citep{bishop1995neural, xia2018instance, aotani2021meta}.

\subsection{Localized Prediction Rule}

For a new observation $x_{\text{new}} \in \mathbb{R}^p$, let $t_{\text{new}} = f(x_{\text{new}})$ and define $\mathcal{J}(x_{\text{new}}) = \{ j : t_{\text{new}} \in I_j \}$ which identify all intervals whose range contains 
$t_{\text{new}}$. For each $j \in \mathcal{J}(x_{\text{new}})$, assign $x_{\text{new}}$ to the nearest cluster
\[
k_j(x_{\text{new}}) = \arg\min_{1 \le k \le K_j} \|x_{\text{new}} - v_{j,k}\|,
\]
where $v_{j,k}$ denotes the centroid of cluster $C_{j,k}$. Define the collection of associated clusters as $\mathcal{C}(x_{\text{new}}) = \{ C_{j,\,k_j(x_{\text{new}})} : j \in \mathcal{J}(x_{\text{new}}) \}$. Because the intervals ${I_j}$ overlap, an observation $x_{\text{new}}$ may belong to one or more intervals and, therefore, may be associated with one or multiple clusters through $\mathcal{C}(x_{\text{new}})$. The primary neighborhood is then defined as the union of these clusters:
\[
U(x_{\text{new}}) = \bigcup_{C \in \mathcal{C}(x_{\text{new}})} C,
\]
Let $U(x_{\text{new}}) = \{u_i\}_{i=1}^m$ denote the resulting set of neighboring observations. This construction ensures that neighborhoods are not defined solely by proximity in the filter space, but also reflect geometric proximity in the original predictor space, thereby adapting to the intrinsic structure of the data. We estimate the conditional class probabilities using a localized weighted average,
\[
\hat{\eta}_r(t_{\text{new}}) = \sum_{i=1}^m \tilde{w}_i \mathbf{1}(y_i = r), \quad r=1,\dots,R.
\]
The weights are defined based on proximity to $x_{\text{new}}$ in the predictor space. Specifically, we use inverse squared distance weighting,
\[
\tilde{w}_i = \frac{\|u_i - x_{\text{new}}\|^{-2}}{\sum_{j=1}^m \|u_j - x_{\text{new}}\|^{-2}}
\]
which assigns greater influence to observations closer to $x_{\text{new}}$ while smoothly downweighting more distant points, providing stronger localization than standard inverse distance weighting while maintaining numerical stability. A few other alternative distance based weighting schemes are also studied in Supplementary Table~S2. For ordinal outcomes, define $\hat{F}(r \mid t_{\text{new}})=\sum_{s=1}^r \hat{\eta}_s(t_{\text{new}})$ and set
$\hat{y}(x_{\text{new}})=\inf\{r:\hat{F}(r \mid t_{\text{new}})\ge 1/2\}. $ We refer to this version as ordinal MAPLE (O-MAPLE). For nominal outcomes, $\hat{y}(x_{\text{new}})=\arg\max_{1\le r\le R}\hat{\eta}_r(t_{\text{new}})$. We refer to this formulation as multinomial MAPLE (M-MAPLE). Thus, O-MAPLE and M-MAPLE share the same localized probability estimator and differ only in the final decision rule used to assign the predicted response category.

\subsection{Theoretical Properties}

The theoretical analysis is conducted by viewing the proposed estimator as a local averaging estimator with data dependent neighborhoods induced by the Mapper graph. Specifically, we analyze the estimator through its effective neighborhood structure, where the filter function restricts observations to subsets with similar projected values in the one dimensional filter space, and clustering within these subsets produces localized groups in the original predictor space $\mathbb{R}^p$. This representation characterizes the estimator as a form of localized averaging, while the effects of clustering and graph connectivity are incorporated through regularity conditions on neighborhood size and diameter. Under these conditions, although the estimator admits a local averaging representation, the neighborhoods are defined through Mapper graph rather than fixed metric proximity. Consequently, the estimator aggregates observations over irregularly shaped neighborhoods induced by graph connectivity and clustering, rather than defining neighborhoods solely through metric distance. This differs from classical nonparametric estimators such as k-nearest neighbor and kernel methods, which rely on fixed metric proximity or bandwidth based smoothing.

\noindent We study the asymptotic behavior of the proposed estimator under the following conditions:

\begin{itemize}
\setlength{\itemindent}{1.5em}
\item[(A1)] $X$ has compact support.
\item[(A2)] The filter function $f$ is Lipschitz continuous.
\item[(A3)] The conditional class probabilities $\eta_r(t)$ are twice continuously differentiable.
\item[(A4)] The Mapper neighborhood size satisfies $|U_n(x)| \to \infty$ as $n \to \infty$.
\item[(A5)] The diameter of clusters in $\mathcal{G}_n$ converges to zero in probability.
\end{itemize}
Details on assumptions, justifications, and proofs of the theoretical results are provided in the Supplementary Section~S2.

\noindent
\textit{
\textbf{Theorem 1} \textit{(Consistency and Bayes risk optimality).}
Suppose assumptions (A1)--(A5) hold. Then, for each class $r = 1, \dots, R$,
\[
\hat{\eta}_r(t) \xrightarrow{P} \eta_r(t), \quad \text{as } n \to \infty.
\]
and the corresponding plug in classifier $\hat{y}(x) = \arg\max_{1 \le r \le R} \hat{\eta}_r(f(x))$ is Bayes risk consistent:
\[
R(\hat{y}) \to R^* \quad (\text{Bayes risk})
\]
}
For the ordinal predictor, the posterior median rule corresponds to an optimal Bayes decision under absolute error loss, so the same convergence result holds \citep{robert2007bayesian}.

\section{Simulation}\label{sec3}

\subsection{Simulation Settings}

We conduct a simulation study to evaluate the predictive performance of MAPLE under nonlinear and heterogeneous data generating mechanisms. The primary objective is to assess its ability to adapt to complex structures, including non-Gaussian covariate distributions and latent branching patterns. Data are generated with three covariates $(x_1, x_2, x_3)$ designed to exhibit skewness and latent branch structure. Specifically, $x_1$ and $x_2$ are sampled from a bivariate log normal distribution.
\[
\log
\begin{pmatrix}
x_1 \\
x_2
\end{pmatrix}
\sim \mathcal{N} \left(
\begin{pmatrix}
0 \\
0
\end{pmatrix},
\begin{pmatrix}
0.25 & 0.09 \\
0.09 & 0.09
\end{pmatrix}
\right)
\]
The third feature $x_3$, was generated using a branch-specific nonlinear transformation. 
$$x_3 = 
\begin{cases}
\cos(1.2 x_1) + 0.6 \log(1 + x_2)^2 + \epsilon, & \text{if } b = 1 \\
\cos(0.6 x_1)\sqrt{x_2} + 0.3 x_2 + \epsilon, & \text{if } b = 2 \\
\tanh(1.2 x_1) + 1.4 \sin(x_2) + \epsilon, & \text{if } b = 3
\end{cases}
\quad ;\quad \text{where} \quad \epsilon \sim \mathcal{N}(0, 0.1^2)
$$
Where observations were assigned to branches $b \in \{1, 2, 3\}$ with probabilities of (0.30, 0.40, 0.30), respectively. The generating scheme of $x_1, x_2$ introduces skewness while $x_3$ mimics a branching pattern, where observations appear to originate from a connected region and then diverge in different directions, similar to the shape of tree branches, without forming three well separated clusters. To achieve this, we combined various trigonometric and nonlinear transformations, including interaction terms, and explored different coefficient settings for each component which explained further in Supplementary Section S5. We considered two scenarios to generate ordinal outcome from the latent continuous variable $\xi$, corresponding to branch specific and global response mechanisms. 

\noindent \textit{Scenario 1:} Branch Specific Latent Scores and Thresholds. In this setting, ordinal responses are generated within each
branch, with the ordering defined locally rather than globally. For each branch $b \in \{1, 2, 3\}$, a latent score $\xi_i$ was defined as a linear combination of predictors $x_1, x_2, x_3$ with branch specific coefficients and random noise $\varepsilon_i$
$$
\xi_i =
\begin{cases}
1.5x_{1i} + 0.04x_{2i} + 0.6x_{3i} + \varepsilon_i, & \text{if } b = 1, \\
0.04x_{1i} + 0.4x_{2i} + 1.5x_{3i} + \varepsilon_i, & \text{if } b = 2, \\
1.0x_{1i} + 0.8x_{2i} + 1.0\log(1 + x_{3i}) + \varepsilon_i, & \text{if } b = 3,
\end{cases}
$$
where $\varepsilon_i \sim N(0, \sigma^2)$.  
To reflect local heterogeneity, branch specific thresholds were computed as quantiles of $\xi_i$ within each branch (25\%, 45\%, 75\%, and 90\%).  
The ordinal response $Y_i$ was then assigned accordingly as given below.
$$
Y_i =
\begin{cases}
\text{Stage 1}, & \xi_i \le q_{0.25}, \\
\text{Stage 2}, & q_{0.25} < \xi_i \le q_{0.45}, \\
\text{Stage 3}, & q_{0.45} < \xi_i \le q_{0.75}, \\
\text{Stage 4}, & q_{0.75} < \xi_i \le q_{0.90}, \\
\text{Stage 5}, & \xi_i > q_{0.90}.
\end{cases}
$$
This setting is designed to induce local heterogeneity in the predictor response relationship, where the same ordinal category corresponds to different latent score ranges across branches, analogous to settings in which a given disease stage may correspond to different biomarker levels across distinct subpopulations. As a result, a single global regression surface is misspecified, and methods that rely on global structure are expected to exhibit higher bias, whereas MAPLE performs localized averaging within graph connected regions, is expected to better adapt to the branching geometry.

\noindent \textit{Scenario 2:} Global Latent Function and Common Thresholds. In the second scenario, a single latent function was used for all observations, irrespective of branch membership.
$$
\xi_i = 0.6x_{1i} + 0.8x_{2i} + 0.6\sin(x_{3i}) + \varepsilon_i
$$
This setting represents a globally smooth signal with a common latent structure across all observations. Because the relationship between predictors and outcome is homogeneous, the advantage of topology adaptive neighborhoods is expected to diminish. In this case, standard regression and tree based methods should perform comparably to the proposed method.

Global thresholds were defined from the quartiles of the entire latent distribution (25\%, 45\%, 75\%, and 90\%), yielding five ordered categories labeled as ``Stage 1'' to ``Stage 5'' following the same rule as above. To examine the impact of data variability and sample size on model performance, we varied both the noise level and the number of observations across simulations. Specifically, the random error term was drawn as $\varepsilon_i \sim N(0, \sigma^2)$ with $\sigma \in \{0.1, 0.2, 0.3\}$, representing low, moderate, and high noise conditions, respectively. Similarly, we evaluated sample sizes of $n \in \{1000, 500, 250\}$ to assess the sensitivity of the methods to decreasing data availability. Together, these scenarios are designed to assess whether the empirical performance of the proposed estimator aligns with the bias-variance and localization arguments developed in Section~2, as reflected in the results presented in Table~\ref{tab1}.

Furthermore, we extended our simulation for scenario 3 with 2 settings to evaluate model robustness under more complex data conditions. In setting 1, we introduced an additional spurious covariate $x_4$ independent of $x_1,x_2,x_3$ such that $Log(x_4) \sim N(0, 0.4^2)$. This variable had no true association with the latent structure or response but was included to assess potential overfitting or false signal detection. In setting 2, we investigated the impact of missing informative predictors by excluding either $x_1$ or $x_2$ while keeping $x_3$ and $x_4$. The variable $x_3$ was intentionally preserved to maintain the inherent branching structure in the data. This configuration reflects realistic scenarios in which a key causal variable is unobserved or an influential confounder remains unmeasured, testing model performance under partial information. For this extension, we fixed the noise level at $\sigma=0.2$ and the sample size at $n=500$, and evaluated the performance under both Scenario 1 and Scenario 2, with results summarized in Table~\ref{tab2}.

Prediction performance is aggregated over 10-fold cross-validation. To ensure valid out of sample evaluation and prevent information leakage, all components of the proposed method are estimated exclusively within the training data in each cross validation split. Specifically, for each fold, the filter function $f$ (via OASDA), the Mapper graph (including clustering), and the cover parameters $(l, S, q)$ are constructed using only the training subset. The resulting fitted Mapper structure is then applied to the held out test data for prediction without any re-estimation. This procedure ensures that no information from test data influences model construction, or tuning, thereby achieving an unbiased assessment of predictive performance. Additional sensitivity analyses evaluating the impact of alternative filter choices on predictive performance are provided in Supplementary Table~S4.

\subsection{Simulation Performance}

\begin{table}[ht]
\centering
\caption{
Out of sample predictive performance under heterogeneous (Scenario 1) and homogeneous (Scenario 2) settings using 10 fold cross validation. Results are reported as mean quadratic weighted kappa ($\kappa_w$) and concordance index (C-index), with empirical standard errors in parentheses. Performance is evaluated across sample sizes ($n = 250, 500, 1000$) and noise levels ($\sigma = 0.1, 0.2, 0.3$), comparing ordinal MAPLE (O-MAPLE), multinomial MAPLE (M-MAPLE), multinomial logistic regression (MLR), ordinal logistic regression (OLR), random forest (RF), and ordinal random forest (O-RF).
}
\vspace{0.3cm}
\label{tab1}
\resizebox{\textwidth}{!}{
\begin{tabular}{|c|c|c|cc|cc|cc|}
\hline
\multirow{2}{*}{\textbf{Scenario}} & 
\multirow{2}{*}{\textbf{Sample}} & 
\multirow{2}{*}{\textbf{Model}} &
\multicolumn{2}{c|}{\textbf{$\sigma$ = 0.1}} & 
\multicolumn{2}{c|}{\textbf{$\sigma$ = 0.2}} & 
\multicolumn{2}{c|}{\textbf{$\sigma$ = 0.3}} \\ 
\cline{4-9}
 & & & 
$\kappa_w$ (SE)& C-index (SE)& 
$\kappa_w$ (SE)& C-index (SE)& 
$\kappa_w$ (SE)& C-index (SE)\\ 
\hline


\multirow{21}{*}{\rotatebox{90}{\textbf{Scenario 1}}}

& \multirow{6}{*}{$n=250$} 
 & O-MAPLE & \textbf{0.868 (0.015)}& \textbf{0.906 (0.007)} & \textbf{0.770 (0.017)}& \textbf{0.854 (0.010)} &  \textbf{0.685 (0.018)}&\textbf{ 0.808 (0.007)} \\ 
 &  & M-MAPLE & 0.850 (0.024)& 0.885 (0.014) & \textbf{0.788 (0.022)}& \textbf{0.849 (0.012)} & \textbf{0.688 (0.021)}& \textbf{0.803 (0.009)} \\ 
 &  & MLR & 0.543 (0.047) & 0.725 (0.020) & 0.509 (0.035) & 0.712 (0.014) &  0.481 (0.039) & 0.699 (0.015)  \\ 
 &  & OLR & 0.594 (0.034) &0.739 (0.015)  & 0.539 (0.025) & 0.711 (0.010) &  0.507 (0.026) & 0.697 (0.010) \\ 
 &  & RF & \textbf{0.851 (0.019)} & \textbf{0.895 (0.009)} & 0.745 (0.017) & 0.836 (0.007)& 0.658 (0.026) & 0.791 (0.009) \\ 
 &  & O-RF & 0.803 (0.026) & 0.871 (0.013) & 0.718 (0.039) & 0.822 (0.015)  & 0.654 (0.023) & 0.783 (0.009)  \\ 
\cline{2-9}

& \multirow{6}{*}{$n=500$} 
 & O-MAPLE &  \textbf{0.847 (0.004)}& \textbf{0.890 (0.002)}& \textbf{ 0.785 (0.009)}& \textbf{0.848 (0.006)} &  \textbf{0.701 (0.010)}& \textbf{0.807 (0.006)} \\ 
 &  & M-MAPLE &  \textbf{0.858 (0.006)}& \textbf{0.891 (0.004)}  &  \textbf{0.772 (0.011)}& \textbf{0.837 (0.006)}  &  0.669 (0.012)& 0.783 (0.007) \\ 
 &  & MLR &  0.498 (0.029) & 0.703 (0.013) &  0.457 (0.029) & 0.686 (0.013) &  0.443 (0.029) & 0.676 (0.013)  \\ 
 &  & OLR & 0.553 (0.021) & 0.722 (0.009) &  0.512 (0.026) & 0.704 (0.011)  &  0.485 (0.033) & 0.694 (0.014) \\ 
 &  & RF & 0.802 (0.012) & 0.874 (0.004) &  0.728 (0.016) & 0.821 (0.007)  & 0.673 (0.016) & 0.791 (0.008) \\ 
 &  & O-RF &  0.815 (0.013) & 0.878 (0.005)  &  0.730 (0.015) & 0.817 (0.008)  & \textbf{0.677 (0.019)} & \textbf{0.789 (0.009)} \\ 
\cline{2-9}

& \multirow{6}{*}{$n=1000$} 
 & O-MAPLE & \textbf{0.881 (0.003)}& \textbf{0.908 (0.002)}  &  \textbf{0.803 (0.007)}&  \textbf{0.859 (0.004)} &\textbf{ 0.735 (0.009)}& \textbf{0.821 (0.005)}  \\ 
 &  & M-MAPLE & 0.874 (0.006)& 0.900 (0.003)  &  0.788 (0.009)& 0.844 (0.005)  & 0.700 (0.011)& 0.797 (0.006)  \\ 
 &  & MLR & 0.604 (0.014) & 0.746 (0.006) &  0.555 (0.020) & 0.720 (0.008)& 0.514 (0.018) & 0.700 (0.008)  \\ 
 &  & OLR & 0.656 (0.011) & 0.771 (0.006)  & 0.621 (0.013) & 0.750 (0.006) &  0.569 (0.015) & 0.725 (0.007)  \\ 
 &  & RF &   \textbf{0.885 (0.005)} & \textbf{0.909 (0.003)}  &  \textbf{0.798 (0.013)} &  \textbf{0.852 (0.008)}  & \textbf{0.713 (0.011)} & \textbf{0.805 (0.006)} \\ 
 &  & O-RF & 0.859 (0.008) & 0.898 (0.005)  &  0.782 (0.010) & 0.844 (0.005)  &  0.710 (0.014) & 0.801 (0.007) \\ 
\hline


\multirow{21}{*}{\rotatebox{90}{\textbf{Scenario 2}}}

& \multirow{6}{*}{$n=250$} 
 & O-MAPLE & \textbf{0.892 (0.005)}& \textbf{0.919 (0.004)}  & 0.802 (0.012)& 0.865 (0.007)  & 0.667 (0.029)& 0.798 (0.014)  \\ 
 &  & M-MAPLE & 0.886 (0.003)& 0.914 (0.004)  & 0.785 (0.014)& 0.855 (0.008)  & 0.652 (0.028)& 0.787 (0.014)  \\ 
 &  & MLR & 0.880 (0.007) & 0.914 (0.005)  & 0.809 (0.011) & 0.870 (0.005)  & 0.660 (0.018) & 0.785 (0.009)  \\ 
 &  & OLR & 0.888 (0.008) & 0.916 (0.006) & \textbf{0.821 (0.010)} & \textbf{0.879 (0.006)}  & \textbf{0.690 (0.019)} & \textbf{0.800 (0.010)}  \\ 
 &  & RF & 0.887 (0.006) & 0.915 (0.005)  & \textbf{0.810 (0.013)} & \textbf{0.871 (0.006)}  & \textbf{0.686 (0.021)} & \textbf{0.800 (0.011)} \\ 
 &  & O-RF & \textbf{0.891 (0.006)} & \textbf{0.918 (0.005)}  & 0.808 (0.013) & 0.869 (0.007)  & 0.684 (0.030) & 0.797 (0.014) \\  
\cline{2-9}

& \multirow{6}{*}{$n=500$} 
 & O-MAPLE & \textbf{0.906 (0.006)}& \textbf{0.921 (0.005)}  & 0.820 (0.010)& \textbf{0.867 (0.006)}  & \textbf{0.732 (0.014)}& \textbf{0.818 (0.008)}  \\ 
 &  & M-MAPLE & 0.900 (0.006)& 0.913 (0.004)  & 0.784 (0.011)& 0.843 (0.006)  & 0.688 (0.017)& 0.791 (0.009)   \\ 
 &  & MLR & \textbf{0.906 (0.005)} & \textbf{0.921 (0.004)}  & \textbf{0.823 (0.010)} &0.860 (0.006)  & 0.726 (0.010) & 0.800 (0.006)   \\ 
 &  & OLR & 0.901 (0.005) & 0.917 (0.004)  & \textbf{0.841 (0.008)} & \textbf{0.876 (0.004)}  & \textbf{0.764 (0.012)} & \textbf{0.829 (0.006)} \\ 
 &  & RF & 0.898 (0.008) & 0.913 (0.005)  & 0.795 (0.013) & 0.849 (0.006)  & 0.687 (0.017) & 0.791 (0.009) \\ 
 &  & O-RF & 0.895 (0.006) & 0.917 (0.004)  & 0.811 (0.008) & 0.863 (0.003)  & 0.697 (0.017) & 0.801 (0.010)  \\  
\cline{2-9}

 & \multirow{6}{*}{$n=1000$} 
 & O-MAPLE & \textbf{0.923 (0.003)}& \textbf{0.935 (0.003)} & \textbf{0.837 (0.006)}& \textbf{0.876 (0.004)}  & \textbf{0.758 (0.008)}& \textbf{0.829 (0.005)} \\ 
 &  & M-MAPLE & \textbf{0.919 (0.004)}& \textbf{0.931 (0.003)}  & 0.815 (0.005)& 0.858 (0.003)  & 0.719 (0.011)& 0.800 (0.005) \\ 
 &  & MLR & 0.901 (0.004) & 0.916 (0.003)  & 0.824 (0.006) & 0.860 (0.003) & 0.739 (0.009) & 0.805 (0.004) \\ 
 &  & OLR & 0.899 (0.004) & 0.917 (0.003)  & 0.827 (0.005) & 0.865 (0.004)  & \textbf{0.747 (0.008)} & \textbf{0.815 (0.004)}  \\ 
 &  & RF & 0.915 (0.004) & 0.928 (0.003)  & 0.812 (0.005) & 0.856 (0.003)  & 0.704 (0.008) & 0.797 (0.004) \\ 
 &  & O-RF & 0.914 (0.003) & 0.929 (0.002)  & \textbf{0.830 (0.008)} & \textbf{0.869 (0.005)}  & 0.731 (0.010) & 0.814 (0.005)  \\ 
\hline

\end{tabular}
}
\end{table}

Table \ref{tab1} summarizes predictive performance across methods, sample sizes, and noise levels, and reveals several consistent patterns. Under Scenario 1, MAPLE consistently outperform global regression models across all settings. For instance, at $n=250$ and $\sigma=0.2$, O-MAPLE achieves $\kappa_w = 0.770$ compared to $0.509$ (MLR) and $0.539$ (OLR), a gap exceeding 0.23, with a similar improvement in C-index (0.854 vs 0.712). This large and persistent discrepancy indicates that global models fail to capture branch dependent relationships, even as sample size increases (e.g., at $n=1000$, $\sigma=0.2$, O-MAPLE: 0.803 vs MLR: 0.555). Relative to tree based methods, MAPLE remains competitive and often superior in smaller samples and higher noise regimes. At $n=250$ and $\sigma=0.3$, O-MAPLE achieves $\kappa_w = 0.685$ compared to $0.658$ for random forest, and exhibits greater stability as noise increases (e.g., at $n=500$, O-MAPLE from $0.847$ to $0.701$ as $\sigma$ increases, whereas RF declines more sharply from $0.802$ to $0.673$). As sample size grows, this gap narrows and RF becomes comparable or slightly better in some low-noise settings (e.g., $n=1000$, $\sigma=0.1$), suggesting that both approaches approximate the underlying structure when sufficient data are available. In contrast, under Scenario 2, the performance differences across methods are minimal. At $n=500$ and $\sigma=0.2$, O-MAPLE achieves $\kappa_w = 0.820$, which is comparable to MLR ($0.823$), OLR ($0.841$), and RF ($0.795$). Similar convergence is observed across all sample sizes, with differences typically within 0.02–0.04. For example, at $n=1000$ and $\sigma=0.1$, all methods achieve $\kappa_w$ values between $0.899$ and $0.923$. This indicates that when the underlying signal is globally homogeneous, the advantage of mapper based localization diminishes, and standard global or tree-based methods perform similarly.

Across both scenarios, increasing noise leads to monotonic performance degradation for all methods. However, the decline is more pronounced for regression models under Scenario 1 (e.g., MLR decreases from $0.543$ to $0.481$ at $n=250$), while MAPLE maintains higher absolute performance levels. The reported standard errors are uniformly small relative to the observed performance gaps, indicating that the differences across methods are stable and not driven by variability across cross-validation folds. Overall, these results show that MAPLE provides the largest gains in structurally heterogeneous settings while remaining competitive in homogeneous regimes, demonstrating its ability to adapt to varying levels of geometric complexity. We further assessed robustness under extreme correlation settings. The results (Supplementary Table S5) indicate that the relative performance of the proposed method remains stable when predictors are either independent or highly correlated.

\begin{table}[ht]
\centering
\caption{
Out of sample predictive performance under Scenario 3. Results are based on 10 fold cross validation and reported as mean quadratic weighted kappa ($\kappa_w$) and concordance index (C-index), with empirical standard errors in parentheses. Setting 1 includes an additional noise covariate ($x_4$), while Setting 2 excludes one informative predictor ($x_1$ or $x_2$). Results are shown for $n=500$ and $\sigma=0.2$, comparing ordinal MAPLE (O-MAPLE), multinomial MAPLE (M-MAPLE), multinomial logistic regression (MLR), ordinal logistic regression (OLR), random forest (RF), and ordinal random forest (O-RF).
}
\vspace{0.3cm}
\label{tab2}
\resizebox{\textwidth}{!}{
\begin{tabular}{|c|c|c|cc|cc|}
\hline
\multirow{2}{*}{\textbf{Covariate Setting}} & 
\multirow{2}{*}{\textbf{Sample}} & 
\multirow{2}{*}{\textbf{Model}} &
\multicolumn{2}{c|}{\textbf{Scenario 1}} & 
\multicolumn{2}{c|}{\textbf{Scenario 2}} \\ 
\cline{4-7}
 & & & $\kappa_w$ (SE)& C-index (SE)& $\kappa_w$ (SE)& C-index (SE)\\ 
\hline

\multirow{7}{*}{\parbox{2cm}{\centering Setting 1\\ \{$x_1,x_2,x_3,x_4$\}}} 
& \multirow{7}{*}{$n=500$} 
 & O-MAPLE & 0.730 (0.010)& \textbf{0.826 (0.006)}   &  \textbf{0.824 (0.006)}&  \textbf{0.872 (0.004)}  \\ 
 & & M-MAPLE & 0.731 (0.011)& 0.812 (0.007) & 0.787 (0.010)&  0.847 (0.006)  \\ 
 &  & MLR & 0.461 (0.025)& 0.686 (0.011) &  0.815 (0.010)&  0.855 (0.006)  \\ 
 &  & OLR &  0.508 (0.024)& 0.702 (0.009) & \textbf{0.841 (0.007)}&  \textbf{0.876 (0.004)} \\ 
 &  & RF & \textbf{0.739 (0.017)}& \textbf{0.825 (0.008)}  & 0.783 (0.012)&  0.845 (0.006)  \\ 
 &  & O-RF & \textbf{0.737 (0.016)}& 0.820 (0.009)  &  0.797 (0.012)&  0.857 (0.006)  \\ 

\hline

\multirow{7}{*}{\parbox{2cm}{\centering Setting 2 (a)\\ \{$x_2,x_3,x_4$\}}} 
& \multirow{7}{*}{$n=500$} 
 & O-MAPLE & 0.623 (0.012)& \textbf{0.776 (0.007)}  & 0.726 (0.011)& \textbf{0.821 (0.005)}   \\ 
 &  & M-MAPLE & 0.582 (0.013)& 0.747 (0.007)   & 0.716 (0.011)& 0.807 (0.005)  \\ 
 &  & MLR & 0.465 (0.025)& 0.683 (0.011)   & 0.713 (0.018)& 0.794 (0.008) \\ 
 &  & OLR &0.496 (0.026)& 0.694 (0.011)  & \textbf{0.734 (0.015)}& 0.811 (0.007) \\ 
 &  & RF & \textbf{0.655 (0.016)}& \textbf{0.781 (0.007)}  & 0.714 (0.012)& 0.807 (0.006) \\ 
 &  & O-RF & \textbf{0.644 (0.016)}& 0.768 (0.008) & \textbf{0.737 (0.012)}& \textbf{0.820 (0.006)}  \\ 

\hline

\multirow{7}{*}{\parbox{2cm}{\centering Setting 2 (b)\\ \{$x_1,x_3,x_4$\}}} 
& \multirow{7}{*}{$n=500$} 
 & O-MAPLE & \textbf{0.573 (0.026)}& \textbf{0.762 (0.010)} & \textbf{0.742 (0.013)}& \textbf{0.828 (0.006)} \\ 
 &  & M-MAPLE & 0.533 (0.025)& 0.738 (0.010) & 0.689 (0.017)& 0.793 (0.008) \\ 
 &  & MLR & 0.202 (0.035)& 0.583 (0.014) & 0.714 (0.014)& 0.798 (0.008)  \\ 
 &  & OLR & 0.304 (0.021)& 0.628 (0.008)  & 0\textbf{.723 (0.015)}& \textbf{0.802 (0.008)}\\ 
 &  & RF & 0.525 (0.019)& 0.736 (0.007) & 0.663 (0.021)& 0.782 (0.010) \\ 
 &  & O-RF & \textbf{0.571 (0.030)}& \textbf{0.748 (0.013)} & 0.687 (0.022)& 0.795 (0.012) \\ 
\hline
\end{tabular}
}
\end{table}

Table \ref{tab2} evaluates robustness of the proposed method under model misspecification, including the presence of irrelevant variables (Setting 1) and omission of informative predictors (Setting 2). When an irrelevant covariate is added (Setting 1), MAPLE maintain performance comparable to the strongest competitors. For example, under Scenario 1, O-MAPLE achieves $\kappa_w = 0.730$, which is close to random forest ($0.739$) and substantially higher than MLR ($0.461$) and OLR ($0.508$). Similar patterns are observed for the C-index, where O-MAPLE attains $0.826$, matching RF ($0.825$). This indicates that the proposed method is not adversely affected by spurious variables and does not overfit to noise. When informative predictors are omitted (Setting 2), performance deteriorates for all methods, but the extent of degradation differs. In Setting 2(a), where $x_1$ is excluded, O-MAPLE achieves $\kappa_w = 0.623$ under Scenario 1, compared to $0.655$ for RF and $0.465$ for MLR, indicating that MAPLE remains competitive despite partial information loss. In Setting 2(b), where $x_2$ (the most influential variable) is excluded, the performance drop is substantial across all methods. However, MAPLE remains among the top performers (O-MAPLE: $0.573$ vs RF: $0.525$ and MLR: $0.202$), highlighting its ability to recover structure from remaining predictors even when key variables are missing.

Under Scenario 2, performance differences across methods are smaller across all settings, consistent with the globally smooth structure. For instance, in Setting 1, all methods achieve similar performance (O-MAPLE: $0.824$, MLR: $0.815$, OLR: $0.841$), and this pattern persists under variable omission, with differences typically within 0.02–0.04. This indicates that when the underlying signal is homogeneous, model misspecification through variable inclusion or exclusion has a more uniform impact across methods. Overall, Table \ref{tab2} shows that MAPLE is robust to both irrelevant variables and partial omission of informative predictors. While all methods degrade when key variables are missing, MAPLE maintains relatively stable performance and avoids the severe deterioration observed in global regression models, particularly under heterogeneous structure.  

\subsection{Interpretation of Mapper Graph}
 
\begin{figure}[h!]
\centering
\resizebox{\textwidth}{!}{

\begin{tabular}{@{}c@{}c@{}}  

\fbox{
\setlength{\tabcolsep}{-3pt} 
\begin{tabular}{@{}l@{}}
\hspace{-5pt}%
 \ \raisebox{-0.07cm}{\textcolor{vividgreen}{\rule{0.4cm}{0.4cm}}}  \hspace{.05cm} \large Stage 1 \\[9pt]
\hspace{-5pt}%
 \ \raisebox{-0.07cm}{\textcolor{lightgreen}{\rule{0.4cm}{0.4cm}}} \hspace{.05cm} \large Stage 2 \\[9pt]
\hspace{-5pt}%
 \ \raisebox{-0.07cm}{\textcolor{yellowish}{\rule{0.4cm}{0.4cm}}} \hspace{.05cm} \large Stage 3 \\[9pt]
\hspace{-5pt}%
  \ \raisebox{-0.07cm}{\textcolor{redish}{\rule{0.4cm}{0.4cm}}} \hspace{.05cm} \large Stage 4 \\[9pt]
\hspace{-5pt}%
 \ \raisebox{-0.07cm}{\textcolor{darkredish}{\rule{0.4cm}{0.4cm}}} \hspace{.05cm} \large Stage 5

\end{tabular}
}

&

\begin{tabular}{ccc}

\large Mapper & \large PCA &\\

\includegraphics[width=0.4\textwidth, height=5cm]{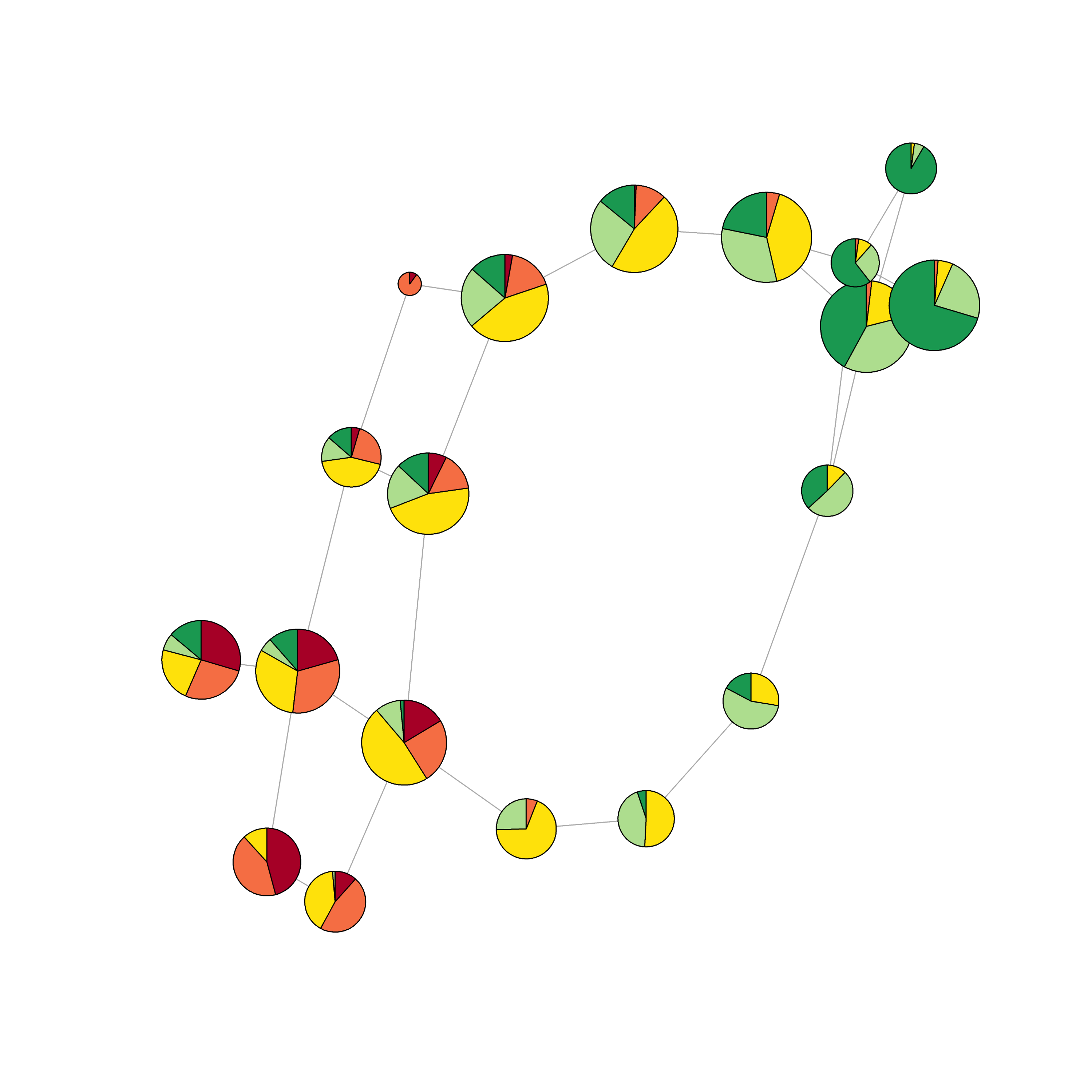} & \includegraphics[width=0.4\textwidth, height=5cm]{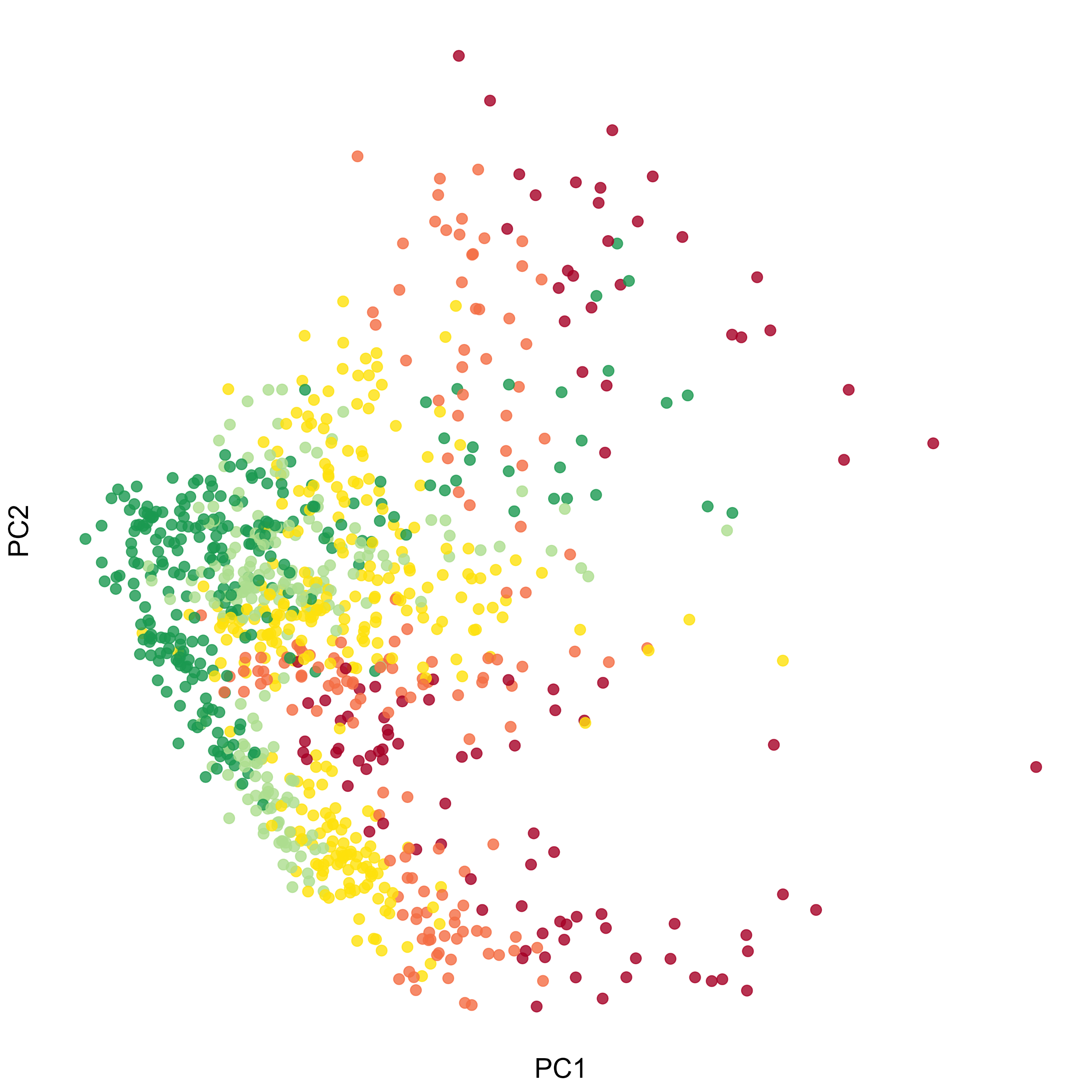} &\raisebox{2.5cm}{\rotatebox[origin=c]{90}{\parbox{2.5cm}{\centering \large Scenario 1}}}\\[6pt]

\includegraphics[width=0.4\textwidth, height=5cm]{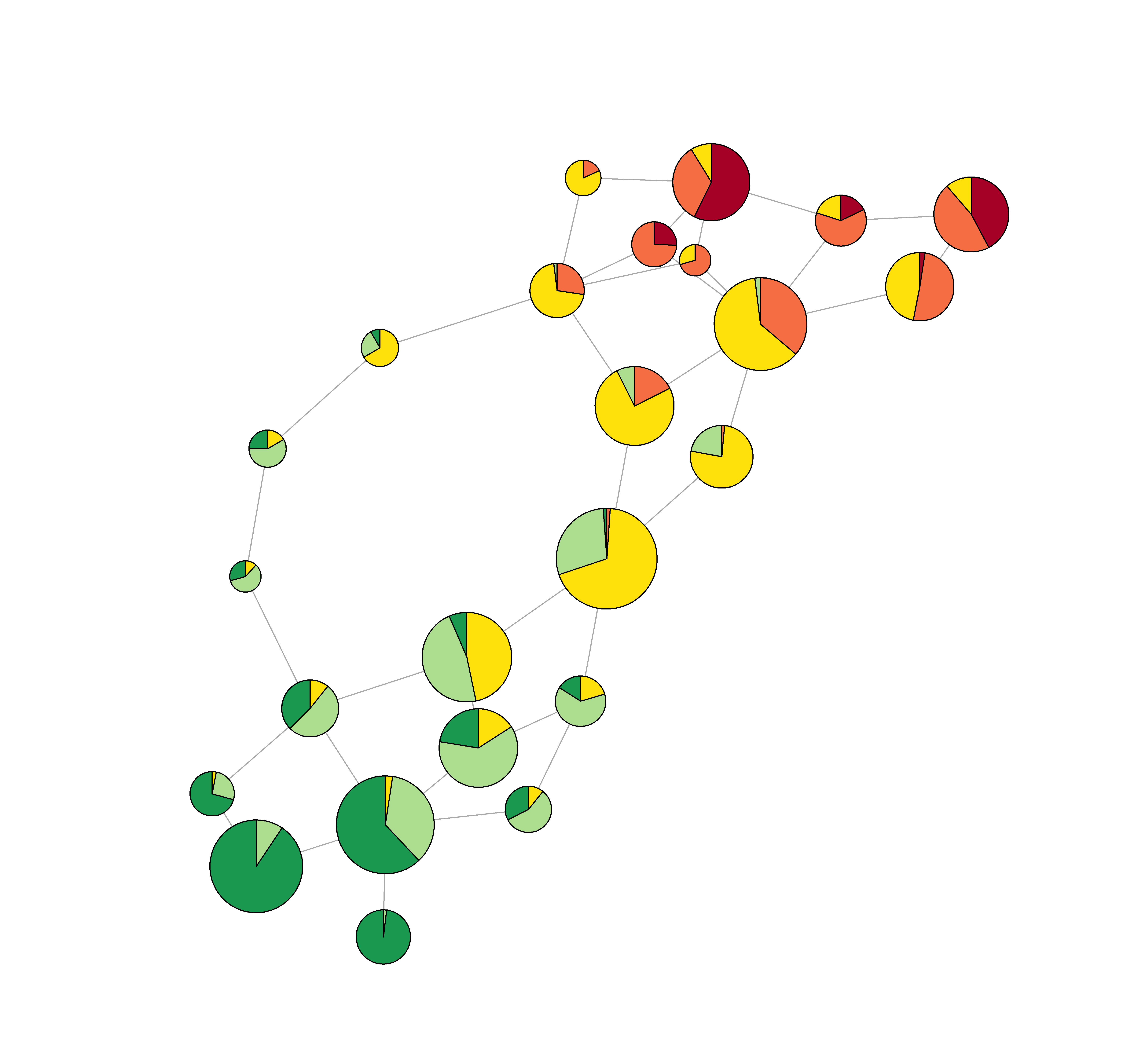} & \includegraphics[width=0.4\textwidth, height=5cm]{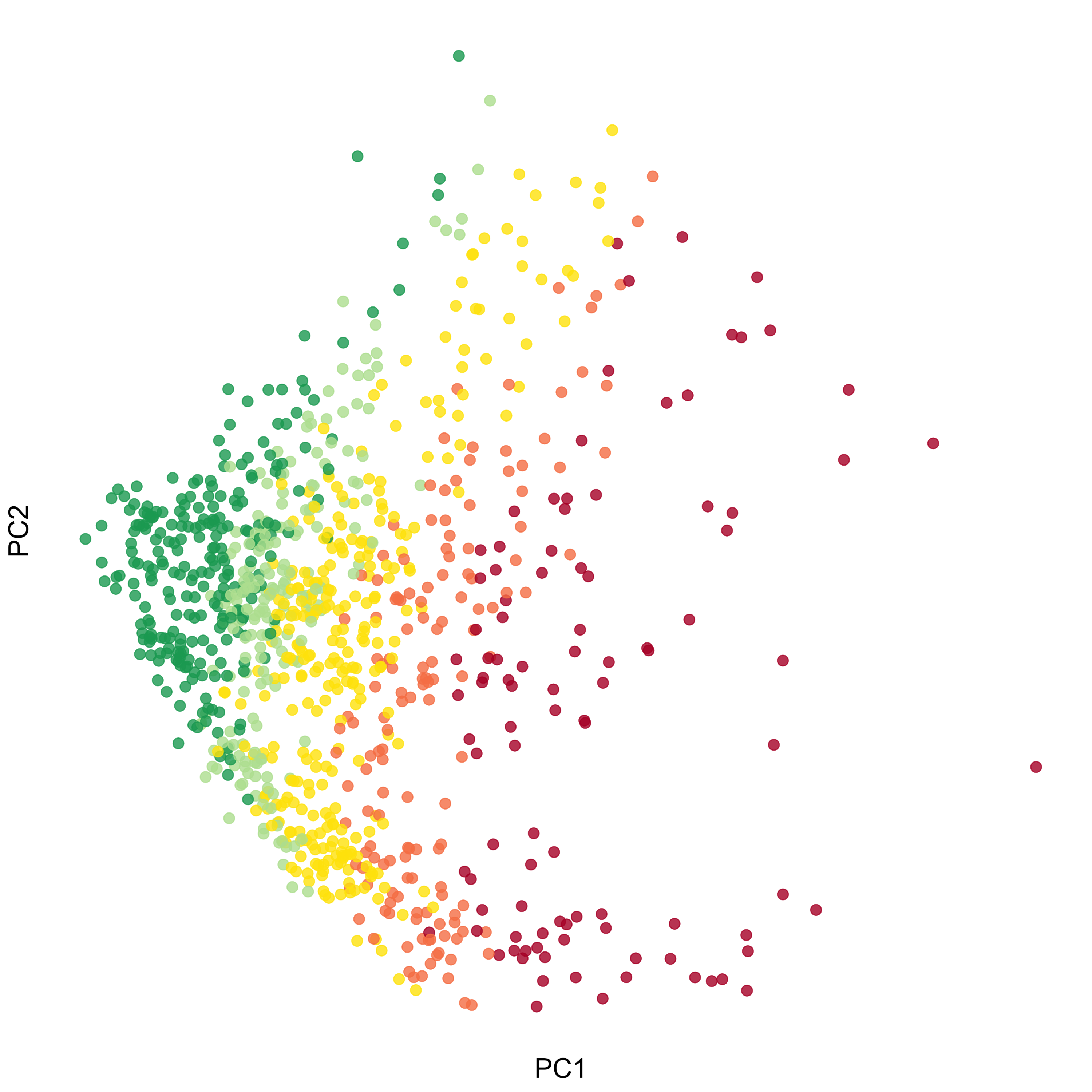} &\raisebox{2.5cm}{\rotatebox[origin=c]{90}{\parbox{2.5cm}{\centering \large Scenario 2}}}\\[6pt]

\end{tabular}

\end{tabular}
}
\caption{
Mapper and PCA representations under Scenario 1 and Scenario 2 for $n=1000$ and $\sigma=0.1$. The Mapper construction uses a bias-variance tuning parameter $\rho=0.1$. Node size reflects cluster size; edges indicate shared observations; colors denote outcome stages.
}
\label{fig: Interpretation of mapper plot 5 noise}

\end{figure}

Figure \ref{fig: Interpretation of mapper plot 5 noise} illustrates the effects of differences in the geometry of the underlying data on the prediction. In the PCA panels, observations are projected into a low dimensional Euclidean space, where substantial overlap between outcome stages is evident in both scenarios, making it difficult to distinguish structural heterogeneity or define meaningful neighborhoods based solely on proximity. In contrast, the Mapper graphs provide a topology aware representation in which each node corresponds to a cluster of observations (with size proportional to sample count), edges indicate shared observations due to overlapping intervals, and pie charts display the distribution of outcome stages within each node. In Scenario 1, the Mapper graph reveals clear branch dependent heterogeneity, with nodes along different connected paths exhibiting distinct and mixed stage compositions, reflecting that the same ordinal stage arises from different latent score regimes across branches. This indicates that the predictor response relationship varies locally across the data manifold, violating the assumptions of a single global regression surface and motivating localized, graph based prediction. In Scenario 2, by contrast, the Mapper graph shows a more globally coherent structure, with stage composition evolving more smoothly across connected nodes, consistent with a single latent function and common thresholds. In this setting, the reduced heterogeneity implies that global models are less misspecified, aligning with the observed convergence in predictive performance. Together, the contrast between PCA and Mapper highlights that while PCA captures variance, it fails to recover the connectivity and local structure that Mapper exploits for improved prediction in heterogeneous settings.

\section{Real Data Applications}\label{sec4}

\subsection{PPMI Data} \label{subsec:ppmi}
 
Parkinson’s disease (PD) is a progressive neurodegenerative disorder characterized by substantial clinical heterogeneity in severity and progression of symptoms. Among the many clinical measures developed to characterize Parkinson’s disease, the Hoehn \& Yahr (H\&Y) scale has remained the most commonly used tool for assessing disease severity since its introduction in 1967. A recent study evaluating the responsiveness of various outcome measures reported that the H\&Y scale was the most sensitive indicator of PD progression. The H\&Y scale provides a straightforward staging system that evaluates overall Parkinsonism impairment, focusing on the extent of bilateral motor involvement and the degree of gait and postural instability \citep{zhao2010progression}. Thus, for the real world application, we considered the reclassified Hoehn \& Yahr Stage (participant is either "OFF" PD medication or untreated for PD) as our ordinal response. We collected baseline data from Parkinson’s Progression Markers Initiative (PPMI), involving 35 clinical sites. Participants were enrolled between  January 2011 and December 2015. The study sample includes 673 patients, among them 235 diagnosed with stage 0 (healthy), 187 with stage 1, and 251 with stage $\geq$2; due to the small number of observations in higher stages, categories $\geq$2 were combined into a single group to ensure stable estimation. Our goal was to evaluate the predictive performance of various clinical, imaging, and CSF biological markers in distinguishing between these stages at baseline. For clinical assessment scores, we included the Movement Disorder Society Unified Parkinson’s Disease Rating Scale (MDS-UPDRS I, II) motor scores, the HVLT Delayed Recall, the Montreal Cognitive Assessment (MoCA) adjusted for education, the University of Pennsylvania Smell Identification Test (UPSIT), geriatric depression scale score (GDS), State-Trait Anxiety Index (STAI) Total Score, Epworth Sleepiness Scale Score (ESS), REM Sleep Behavior Disorder Screening Questionnaire (RBDSQ) total score, Modified Schwab \& England Activities of Daily Living Score (ADL). Also, we added cerebrospinal fluid (CSF) biomarkers, specifically amyloid-beta 1-42 (A$\beta$42), alpha-synuclein ($\alpha$-syn), total tau (t-tau), and Phosphorylated tau (p-tau 181).

In practice, a preliminary variable screening step was performed to reduce dimensionality and mitigate the influence of weakly informative predictors in the high dimensional setting. Predictive performance was evaluated using the quadratic weighted kappa (QWK), which accounts for the ordinal nature of the response \citep{cohen1968weighted},
\[
\mathrm{QWK}:=\kappa_w
=
1-\frac{\sum_{i,j} w_{ij}O_{ij}}{\sum_{i,j} w_{ij}E_{ij}},
\qquad
\Delta^{(j)}
=
100\times
\frac{
\mathrm{QWK}_{\mathrm{original}}
-
\mathrm{QWK}_{\mathrm{permute}(j)}
}
{\mathrm{QWK}_{\mathrm{original}}},
\]
where \(O_{ij}\) and \(E_{ij}\) denote the observed and expected agreement matrices, respectively, and \(w_{ij}=(i-j)^2\). Variable importance was quantified through the relative reduction in predictive performance after permutation of the $j$-th covariate. Since MAPLE predictions are constructed from Mapper-based local neighborhoods, the resulting importance measure reflects both predictive contribution and topological neighborhood structure. Importance scores are computed across cross validation folds and summarized using the median importance score $\tilde{\Delta}^{(j)}$. Variables with larger values of $\tilde{\Delta}^{(j)}$ were considered more influential for prediction. This approach is Conceptually related to permutation importance used in ensemble methods \citep{strobl2008conditional}. Additional simulation results evaluating the variable screening procedure are provided in Supplementary Section S3.

After variable selection, the model retained four predictors exceeding a 5\% threshold: MDS-UPDRS Part II score ($\tilde{\Delta}=92.13$), UPSIT ($\tilde{\Delta}=26.36$), RBDSQ/REM-related score ($\tilde{\Delta}=5.41$), and STAI total score ($\tilde{\Delta}=5.31$), indicating that most of the predictive signal was concentrated in a small set of clinically interpretable markers spanning motor impairment, olfactory dysfunction, sleep-related symptoms, and anxiety burden . The predictive performance of mapper and competitive methods are shown in Table \ref{tab:real_performance}.

\begin{figure}[ht]
\centering
\resizebox{\textwidth}{!}{

\begin{tabular}{@{}c@{}c@{}}  

\fbox{
\setlength{\tabcolsep}{-3pt} 
\begin{tabular}{l}
\hspace{-5pt}%
 \ \raisebox{-0.07cm}{\textcolor{vividgreen}{\rule{0.4cm}{0.4cm}}}  \hspace{.05cm} \large Stage 0 \\[9pt]
\hspace{-5pt}%
 \ \raisebox{-0.07cm}{\textcolor{yellowish}{\rule{0.4cm}{0.4cm}}} \hspace{.05cm} \large Stage 1 \\[9pt]
\hspace{-5pt}%
 \ \raisebox{-0.07cm}{\textcolor{darkredish}{\rule{0.4cm}{0.4cm}}} \hspace{.05cm} \large Stage $\geq$2

\end{tabular}
}
&

\begin{tabular}{ccc}

\large Mapper & \large t-SNE & \large PCA\\

\includegraphics[width=0.28\textwidth]{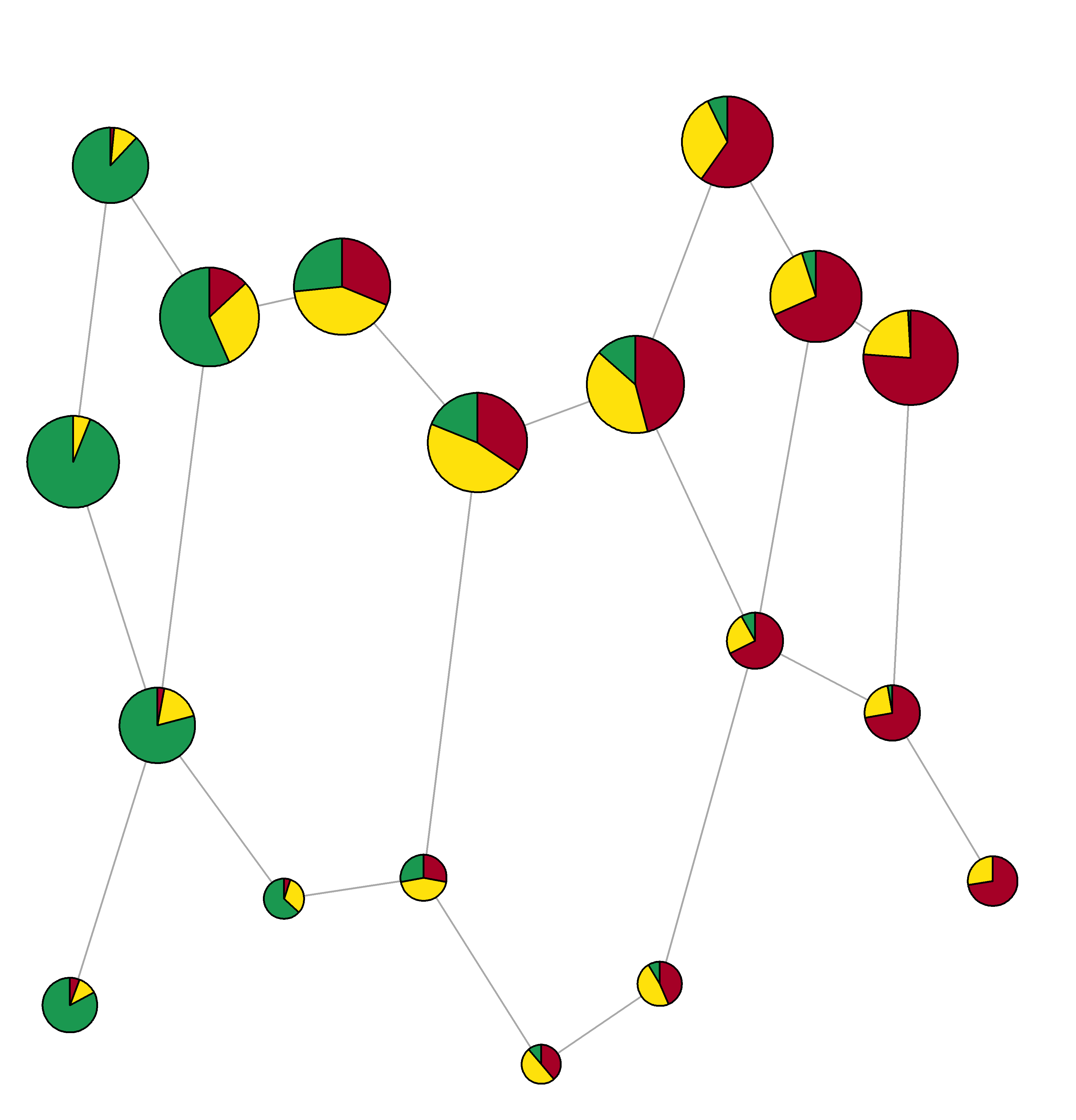} &
\includegraphics[width=0.28\textwidth]{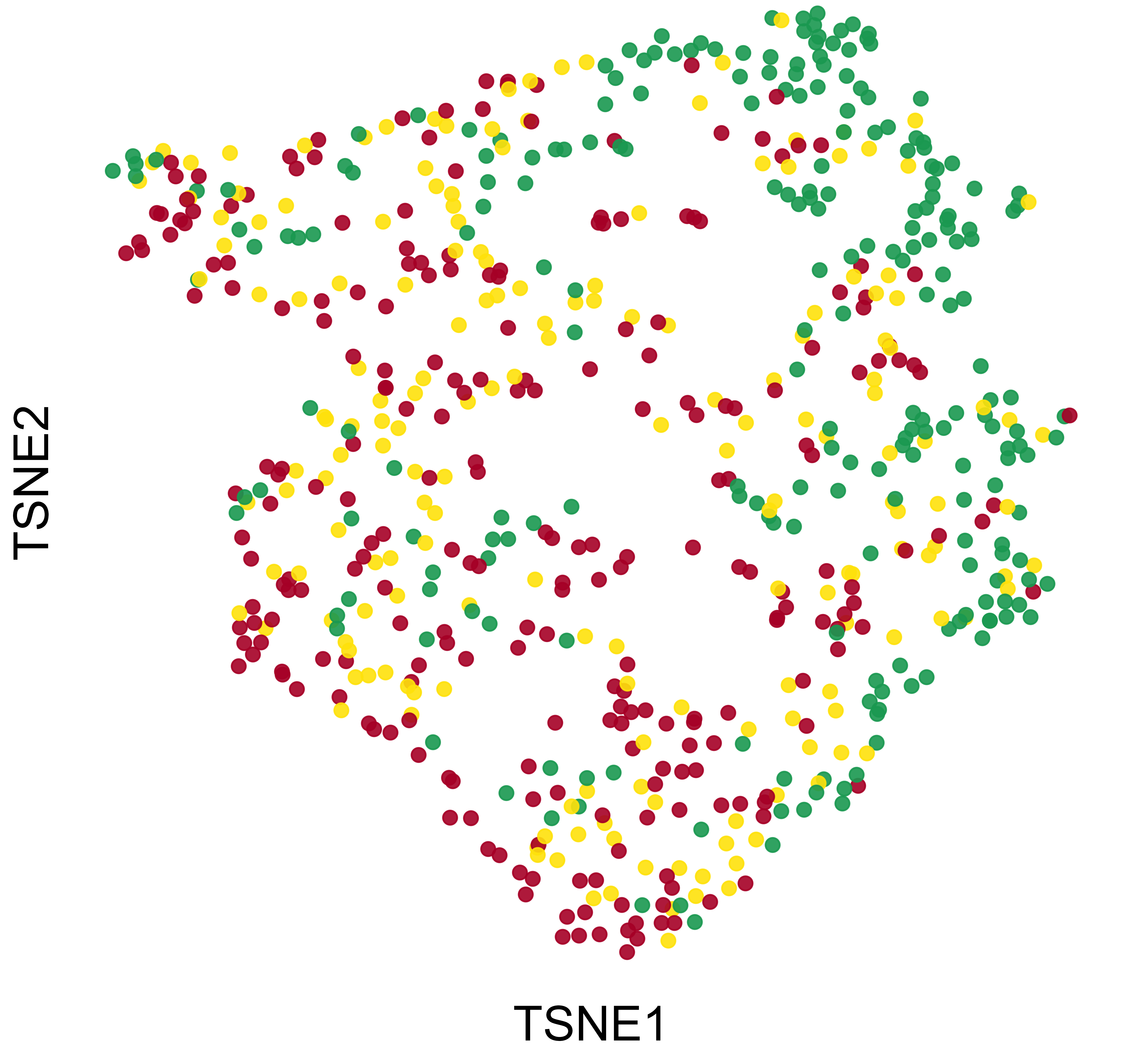} &
\includegraphics[width=0.28\textwidth]{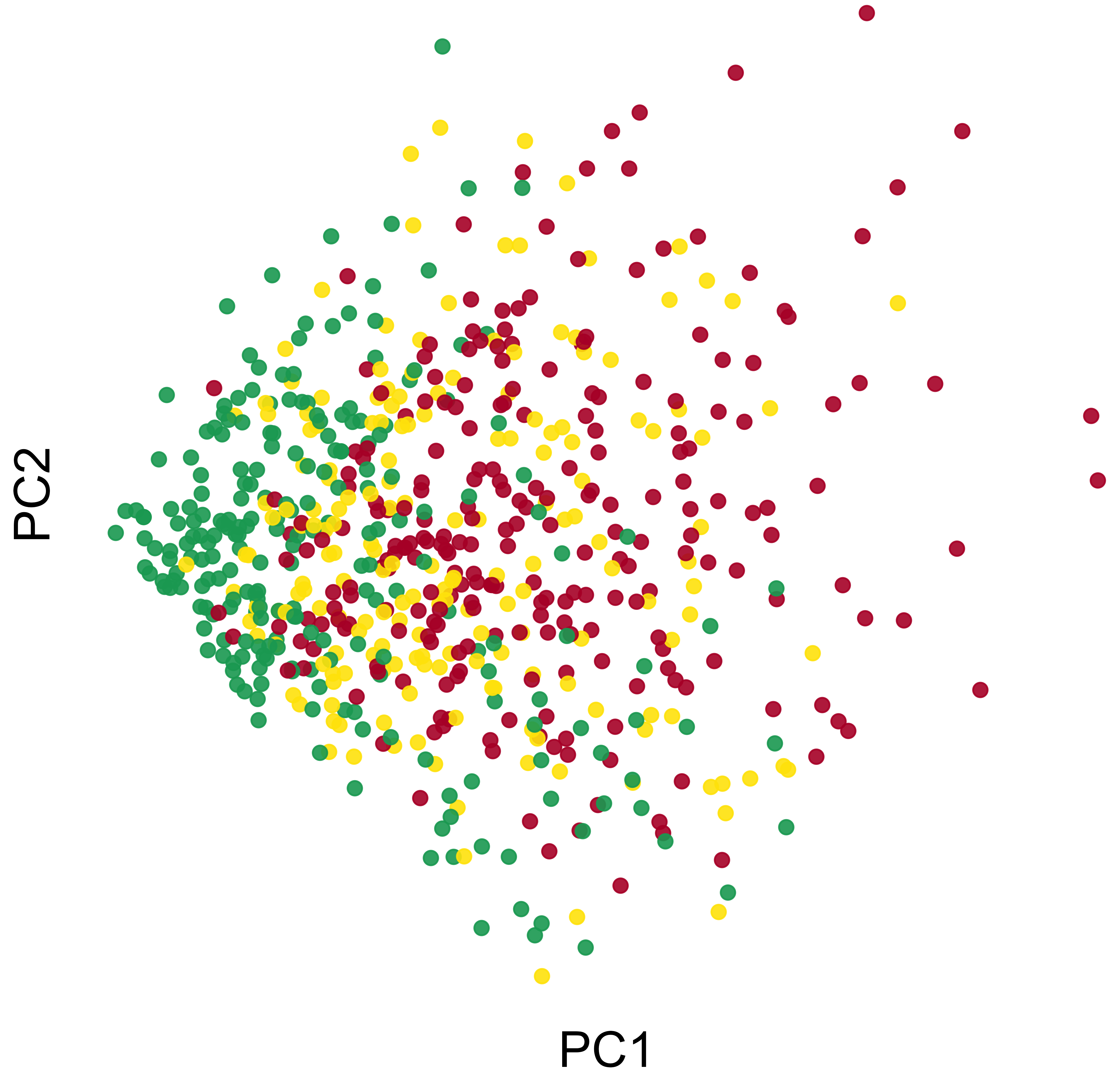}

\end{tabular}

\end{tabular}
}
\caption{
Mapper graph constructed using the bias-variance tuning parameter $\rho = 0.1$ along with t-SNE and principal component analysis projections of the PPMI dataset, colored by Hoehn \& Yahr stages.
}
\label{fig: mapper plot PPMI}
\end{figure}

Figure \ref{fig: mapper plot PPMI} compares topology based and classical dimension reduction approaches to visualize the structure of the PPMI data. The PCA and t-SNE representations project the data into low dimensional Euclidean space, where substantial overlap between disease stages is observed, limiting the ability to distinguish clinically meaningful subgroups or define well separated neighborhoods based solely on proximity. In contrast, the Mapper representation reveals a structured and heterogeneous organization of the data. Nodes along the graph define patient subgroups with distinct disease stage compositions, indicating that individuals within the same Hoehn \& Yahr stage are not clinically homogeneous. This pattern indicates that the relationship between clinical markers and disease stage differs across patient subgroups, suggesting that a single global model may mask subgroup-specific progression patterns. Transitions across connected nodes reflect gradual shifts in stage composition, suggesting that disease progression occurs along multiple continuous pathways rather than a single uniform trajectory. Clinically, this implies that patients assigned to the same Hoehn \& Yahr stage may differ in their underlying progression risk, meaning that stage alone may not fully capture disease severity or future trajectory. For example, a patient classified as Stage 1 but located in a region of the graph enriched with Stage $\ge 2$ cases may represent a subgroup at elevated risk of progression. Global models that assume a uniform relationship between predictors and disease stage may average across these subgroups, potentially leading to biased or less accurate predictions. By contrast, the Mapper representation identifies regions of the patient population with distinct stage compositions, providing a framework for understanding subgroups of patients who may follow different clinical trajectories. By identifying these subgroups, the mapper based approach enables more refined risk stratification and may help identify patients who require closer clinical monitoring despite being classified at an early stage.

\subsection{UCSC Xena Glioma Data}

Glioma is one of the most common primary brain tumors and accounts for greater than 70\% of malignant brain tumors, \cite{cao2021development} associated with the central nervous system are diffuse gliomas with their origins in glial cells \citep{kocaturk2023identification}. Standard treatment, such as mass surgical resection, radiotherapy, and chemotherapy are aggressive, but their prognosis remains unsatisfactory \citep{cao2021development, kocaturk2023identification, lin2025establishment}. Diffuse gliomas are classified into two major subtypes being low grade glioma (LGG, grade II/III) and glioblastoma multiforme (GBM, grade IV). The invasive and therapy resistant nature of GBM plays a fundamental role in its poor prognosis \citep{kocaturk2023identification}. The relationship between gene expression profiles and cancer progression has been documented across numerous studies \citep{kocaturk2023identification}. Thus, we collected data from TCGA lower grade glioma and glioblastoma (GBMLGG) gene expression RNAseq and Phenotypes obtained from UCSC Xena database \citep{goldman2020visualizing}. The dataset included 667 patients with solid tumor, consisting of 249 with grade II glioma, 265 with grade III glioma, and 153 with grade IV glioblastoma. We retained top 10\% high variability gene for analysis consisting of 2053 genes. Following variable selection, 44 genes were selected using random forest variable importance scores (mean decrease in accuracy) with a threshold of 3. This preselection step was implemented due to the computational cost of MAPLE variable selection, as reflected in Supplementary Table S6. Based on the selected genes, the predictive performance of mapper and competitive methods are shown in Table \ref{tab:real_performance}.

\begin{figure}[h!]
\centering
\resizebox{\textwidth}{!}{
\begin{tabular}{c}

\includegraphics[width=1\textwidth, height=12cm]{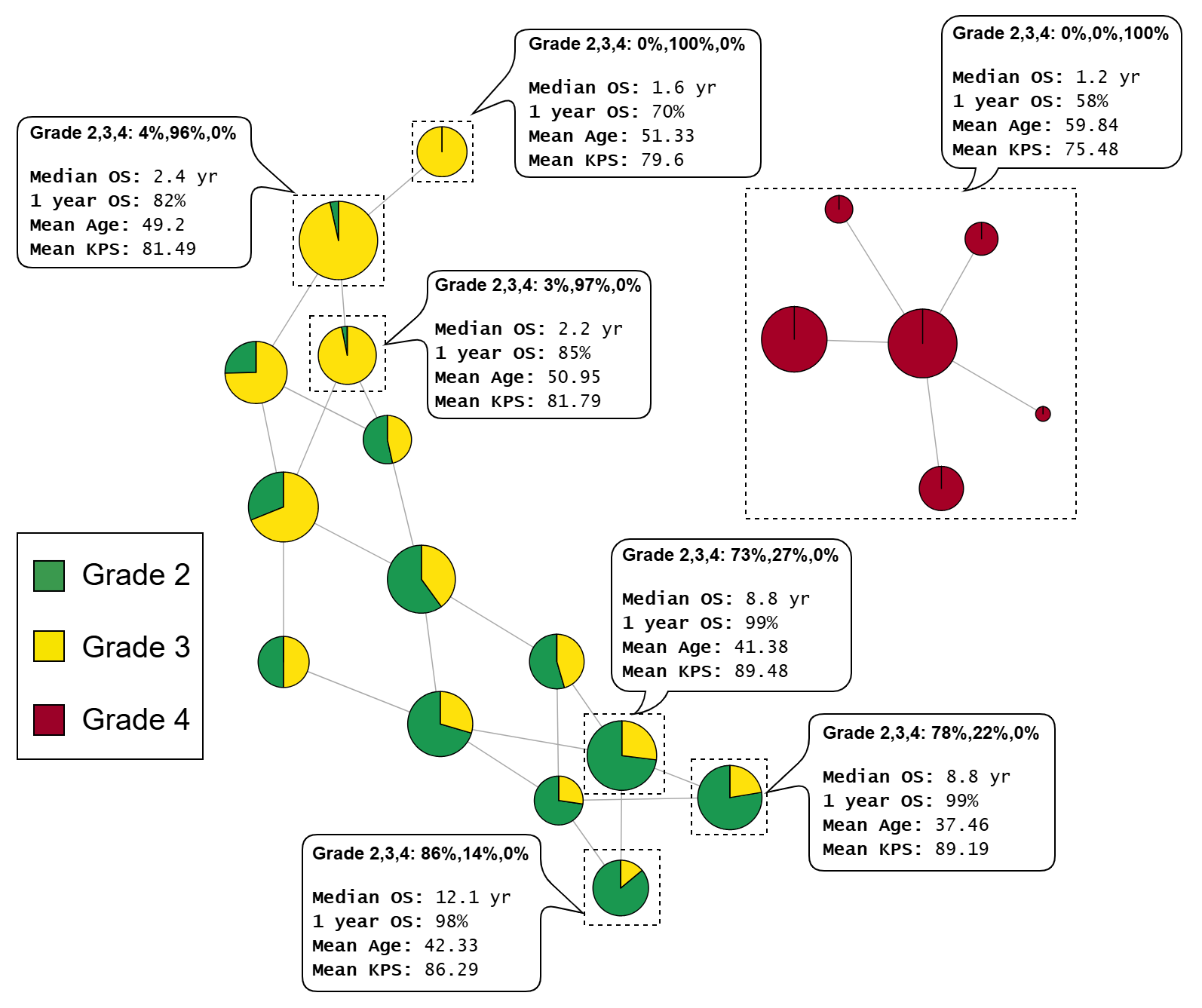} \\[6pt]

\end{tabular}
}
\caption{Mapper graph of the UCSC Xena glioma dataset constructed with tuning parameter $\rho = 0.1$, corresponding to a high resolution (fine granularity) cover. Nodes represent clusters with size proportional to sample count and pie charts indicating tumor grade distribution. Annotated regions highlight representative subnetworks with distinct grade compositions and associated clinical summaries, including median overall survival (OS), 1-year OS, mean age, and mean Karnofsky Performance Status (KPS), illustrating heterogeneous survival patterns.}

\label{Final_UCSC_Xena}
\end{figure}

The Mapper representation of the UCSC Xena glioma dataset in Figure \ref{Final_UCSC_Xena} identifies clinically distinct tumor subgroups. Annotated summaries for Grades 2 and 3 are presented for the top three Mapper nodes by grade proportion, restricted to nodes with sample size at least 40. A dominant component contains Grade 2-3 tumors, while a separate cluster of Grade 4 tumors indicates that high grade glioblastoma forms a distinct disease subgroup rather than a continuation of lower grade progression. Within the larger component, the mixing of Grades 2 and 3 suggests a continuum of disease progression, where intermediate grade tumors may not form clearly separable categories in terms of underlying biology. The Grade 4 cluster shows near complete purity, indicating a high risk subgroup with substantially different clinical behavior compared to lower grade tumors. Node level summaries demonstrate consistent and aligned gradients across multiple clinical variables: overall survival decreases, age increases, and Karnofsky Performance Scale (KPS) declines as one moves from lower grade to higher grade regions of the graph. Specifically, Grade 2 dominant regions are characterized by longer survival, younger age, and higher functional status, whereas the Grade 4 cluster corresponds to considerably shorter survival, older patients, and reduced KPS. This alignment between tumor grade, survival, age, and KPS measures indicates that the identified subgroups correspond to clinically meaningful risk strata rather than arbitrary clusters. For example, patients located near the boundary between Grade 3 and Grade 4 regions may represent a transitional high risk group with worse prognosis than typical Grade 3 cases, suggesting potential value for closer monitoring or more aggressive treatment. These findings demonstrate that tumor severity is not fully captured by grade alone, and that patients within the same histological category may differ in prognosis depending on their position within the data structure. By incorporating this structure, the Mapper based approach provides a framework for identifying clinically relevant subgroups that may improve prognostic assessment beyond standard grading systems. Parallel analyses using PCA and t-SNE similarly separated Grade 4 tumors from lower-grade gliomas, reflecting a strong global signal of tumor severity. However, these projections do not provide region-specific summaries of subgroup composition and therefore do not capture the localized structure evident in the Mapper representation.

\subsection{Comparative out of sample Performance}

\begin{table}[h!]
\centering
\caption{
Comparative out of sample predictive performance for the PPMI and UCSC Xena applications. Results are based on cross validated prediction and reported as mean quadratic weighted kappa ($\kappa_w$) and concordance index (C-index), with empirical standard errors in parentheses. For PPMI, models were fit using the four predictors retained after variable selection. For UCSC Xena, models were fit using the 44 selected genes. Higher values indicate better ordinal agreement and discrimination.
}
\label{tab:real_performance}
\vspace{0.3cm}
\begin{tabular}{|l|c|c|c|c|}
\hline
& \multicolumn{2}{c|}{\textbf{PPMI}} 
& \multicolumn{2}{c|}{\textbf{UCSC Xena}}  \\
\cline{2-3} \cline{4-5}
\textbf{Model} 
& \textbf{$K_w$} (SE) 
& C-index (SE) 
& \textbf{$K_w$} (SE) 
& C-index (SE) \\
\hline
O-MAPLE  &  \textbf{0.735 (0.013)}& \textbf{0.836 (0.007)}  & \textbf{0.835 (0.012)} & \textbf{0.883 (0.010)} \\
M-MAPLE  &  0.690 (0.016)& 0.804 (0.007)  &  \textbf{0.835 (0.012)} & \textbf{0.883 (0.010)}  \\
MLR       &  \textbf{0.697 (0.013)}& \textbf{0.812 (0.007)}  & 0.820 (0.017) & 0.876 (0.013) \\
OLR       &  0.675 (0.021)& 0.809 (0.010)   &  0.803 (0.013) & 0.864 (0.010)   \\
RF        &  0.680 (0.013)& 0.805 (0.007)   &  0.821 (0.014) & 0.874 (0.011)  \\
O-RF      &  0.696 (0.014)& 0.807 (0.007)   & 0.825 (0.012) & 0.876 (0.010) \\
\hline
\end{tabular}
\end{table}

For the PPMI data with the reduced feature set (Table \ref{tab:real_performance}), the ordinal MAPLE estimator achieved the strongest overall performance, with $\kappa_w = 0.735$ (SE 0.013) and C-index = 0.836 (SE 0.007), exceeding multinomial MAPLE, multinomial logistic regression, ordinal logistic regression, random forest, and ordinal random forest. The strongest competitors were multinomial logistic regression for weighted kappa (0.697) and random forest–based methods for discrimination, but all remained below ordinal MAPLE on both metrics. These results suggest that disease stage in the PPMI cohort is not fully captured by a single global model. Instead, the improved performance of ordinal MAPLE is consistent with local heterogeneity in the relationship between biomarkers and Hoehn \& Yahr stages, where graph based neighborhoods provide a more appropriate basis for prediction. This interpretation is supported by the corresponding Mapper visualization, in which stage composition changes gradually across connected regions rather than forming well separated clusters.

For the UCSC Xena glioma data, performance was uniformly high across methods, reflecting the presence of a strong signal for tumor grade in the selected RNA sequencing features. Both ordinal and multinomial MAPLE achieved the highest observed performance, with $\kappa_w = 0.835$ (SE 0.012) and C-index = 0.883 (SE 0.010), narrowly exceeding logistic regression and random forest approaches. Although the gains over the strongest competitors were smaller than in PPMI, the results still indicate that MAPLE remains fully competitive in a high dimensional genomic setting while preserving interpretability through its graph representation.

\section{Discussion}\label{sec5}

This article presents the MAPLE algorithm, a robust extension of the Mapper framework for predictive modeling in high dimensional settings. By formulating Mapper based prediction as a nonparametric local averaging estimator with data adaptive, graph induced neighborhoods, the proposed approach integrates data driven geometric structure into the estimation of conditional class probabilities. In addition, the MAPLE algorithm incorporates a data driven procedure for selecting cover parameters based on a bias-variance trade off, extends naturally to ordinal and nominal outcomes, and provides a permutation based variable importance measure for assessing covariate contributions. Theoretical results establish pointwise consistency and Bayes risk optimality under standard regularity conditions, while simulation studies and real data applications demonstrate that the method achieves strong predictive performance, particularly in settings characterized by nonlinear, heterogeneous, or structured relationships where common parametric models are likely to be misspecified. When data exhibits local heterogeneity tied to latent structure, ignoring that structure can lead to biased predictions regardless of model complexity. The role of Mapper is not simply to provide a new prediction algorithm but to alter how neighborhoods are defined. By constructing neighborhoods that follow connectivity in the data rather than coordinate directions, the method shifts the focus from approximating functions globally to respecting the geometry on which those functions are defined.

At the same time, the proposed method is not universally advantageous. When the underlying relationship between predictors and response is globally smooth and well approximated by parametric or tree based models, the additional flexibility introduced by the Mapper construction may yield limited gains and can even increase variance. In such cases, standard approaches may achieve comparable or superior performance due to lower estimation variability. In addition, the method depends critically on the choice of filter function, which governs how the high dimensional data are projected into a lower dimensional representation. If the filter fails to preserve relevant structural features, the resulting Mapper graph may not accurately reflect the intrinsic geometry of the data, leading to suboptimal neighborhood construction.

A further practical consideration is computational cost. Constructing the Mapper graph, particularly within a cross validation framework and under repeated permutations for variable importance, can be computationally intensive in moderate to large datasets. As shown in Supplementary Table S6, runtime increases with dimensionality due to repeated prediction steps. However, this computational burden is primarily incurred during the model selection phase, especially in the evaluation of variable importance. In this step, conservative thresholds (e.g., 3–5\%) provide a practical guideline for identifying uninformative covariates, although these should be interpreted relative to overall model performance: when predictive accuracy is already high, small incremental gains may not justify retaining additional variables, whereas in moderate performance settings even modest improvements may be meaningful. Once a stable subset of covariates is identified, subsequent analyses can be conducted on a reduced feature space, substantially mitigating computational demands. In this sense, the cost represents an upfront investment that yields a more parsimonious and interpretable model. Future work will focus on developing an integrated variable selection strategy within the MAPLE framework, in which predictor selection is incorporated directly into the model construction process rather than performed as a separate preliminary screening step. Such an approach would allow the topological representation and prediction procedure to adapt simultaneously to the most informative features, potentially improving both computational efficiency and predictive performance in high dimensional settings.

An important feature of the proposed framework is the way neighborhoods are defined through the Mapper graph. In this paper, we focus on primary neighborhoods, constructed from clusters directly associated with a new observation. This choice provides a balance between locality and stability in estimation. However, the Mapper framework naturally allows for extensions beyond this definition. In particular, one may consider secondary neighborhoods, defined by including clusters that are directly connected to the primary nodes through an immediate layer of graph connectivity. Such extended neighborhoods capture broader structural relationships and increase the effective neighborhood size, which may influence the variability of the estimator. However, as illustrated in Supplementary Table S3, this extension does not yield consistent improvements in predictive performance relative to the primary neighborhood definition. This highlights a key feature of Mapper based methods, that is locality can be defined through graph topology rather than purely through metric distance. A systematic investigation of the depth of the neighborhood and its impact on prediction remains an interesting direction for future work.

The results of this paper indicate a potential shift in high dimensional prediction, in which capturing the shape of the data serves as a viable alternative to generalizing models to accommodate complexity. Several extensions naturally follow from this framework. First, many biomedical datasets involve mixed data types, and extending Mapper based neighborhood construction beyond continuous predictors is an important direction. Second, adapting the method to time-to-event outcomes would allow it to be applied in survival analysis, where heterogeneity and nonlinear structure are common but difficult to model with standard approaches.

\section*{Disclosure statement}
Authors declare no conflicts of interest

\if1\blind
{
\acks{PD and NM are partially supported by the National Institutes of Health/National Cancer Institute Cancer Center Support Grant P30 CA016059.}
}
\fi

\section*{Code and Data}

\if1\blind
{
Code for MAPLE, including demonstration scripts, is available on GitHub at \url{https://github.com/MoinulAhsan/MAPLE}. Parkinson’s disease data were obtained from the Parkinson’s Progression Markers Initiative (PPMI; \url{https://www.ppmi-info.org}). Access to these data requires registration and approval through the PPMI data portal. All analyses were conducted in accordance with the PPMI data use agreement. Glioma RNA sequencing data were obtained from The Cancer Genome Atlas (TCGA) via the UCSC Xena platform (\url{https://xena.ucsc.edu/}), which provides publicly available cancer genomics data.

\fi
\if0\blind
{
Code for MAPLE, including demonstration scripts, will be made available on GitHub. Parkinson’s disease data were obtained from the Parkinson’s Progression Markers Initiative (PPMI) (\url{https://www.ppmi-info.org}). Access to these data requires registration and approval through the PPMI data portal. All analyses were conducted in accordance with the PPMI data use agreement. Glioma RNA sequencing data were obtained from The Cancer Genome Atlas (TCGA) via the UCSC Xena platform (\url{https://xena.ucsc.edu/}), which provides publicly available cancer genomics data.

}
\fi







\vskip 0.2in
\clearpage
\bibliographystyle{plainnat}
\bibliography{wileyNJD-AMA}

\end{document}